\documentclass[twocolumn,showpacs,preprintnumbers,amsmath,amssymb]{revtex4}

\usepackage{graphicx}
\usepackage{dcolumn}
\usepackage{bm}
\newcommand{\be}{\begin{equation}}
\newcommand{\ee}{\end{equation}}
\newcommand{\bef}{\begin{figure}}
\newcommand{\eef}{\end{figure}}
\newcommand{\bea}{\begin{eqnarray}}
\newcommand{\eea}{\end{eqnarray}}
\newcommand{\bx}{{\bf x}}
\newcommand{\by}{{\bf y}}
\newcommand{\bu}{{\bf u}}
\newcommand{\bv}{{\bf v}}
\newcommand{\bk}{{\bf k}}
\newcommand{\bq}{{\bf q}}
\newcommand{\bH}{{\bf H}}

\begin{document}

\preprint{APS/123-QED}


\title{Point processes and stochastic displacement fields}


\author{Andrea Gabrielli$^{1}$}
\affiliation{$^{1}$ Statistical Mechanics and Complexity Center - INFM, 
Department of Physics, University ``La Sapienza'' of Rome, Piazzale
Aldo Moro 2, 00185-Rome, Italy.}

\date{\today}
\begin{abstract}
The effect of a stochastic displacement field on a statistically
independent point process is analyzed.  Stochastic displacement fields
can be divided into two large classes: spatially correlated and
uncorrelated.  For both cases exact transformation equations for the
two-point correlation function and the power spectrum of the point
process are found, and a detailed study of them with important
paradigmatic examples is done.  The results are general and in any dimension. 
A particular attention is devoted to the kind
of large scale correlations that can be introduced by the displacement
field, and to the realizability of arbitrary ``superhomogeneous'' point
processes.
 
\end{abstract}

\maketitle

\section{Introduction}

Point processes (i.e., stochastic spatial distributions of
point-particles with identical mass) are very useful mathematical
models of many $n-$body and complex systems. Crystals (regular,
perturbed, and/or defected)
\cite{ziman,ashcroft,groma}, quasi-crystals \cite{radin}, structural
glasses, fluids \cite{hansen}, cosmological self-gravitating systems
\cite{pee93,n-body1}, and also computer image processing problems \cite{cpu},
and bio-metrical studies \cite{renshaw} are only some examples of
systems which are usually represented as specific point processes with
appropriate spatial correlation properties.

The study of this branch of stochastic processes and the discovery
of new statistical properties can be of fundamental importance in many
scientific topics. 
Many mathematical studies have been already done about
this class of processes and many useful results have been derived
(e.g., see \cite{daley,kertscher,torquato98}).

One important question about a point process is what happens to its
statistical properties when it is perturbed by a stochastic spatial
deformation that can have in turn an internal degree of spatial
organization, that is, spatial correlations. Depending on the physical
application and the context, the perturbation can be seen either as a
fluctuation due to a physical process or as noise.  The fundamental
question consists in finding how the spatial correlation properties of
the point process change under the effect of the perturbation, and how
effective this can be in changing the spatial correlations of the 
point process.

In this paper we focus our attention on the changes induced on the
two-point spatial correlations of a point process by a stochastic
displacement field both with and without displacement-displacement
correlations. We work in the hypothesis of statistical independence
between the point process and the displacement field.  The exact
results presented in this paper can find application in many
scientific topics. For instance, in the context of the so-called
$n-$body cosmological simulations \cite{n-body1}, performed to study
the problem of ``structure formation'' (e.g., galaxy formation) from
the primordial matter density field under the effect of the internal
gravitational interaction, point processes are used to represent the
evolving matter density field.  The initial conditions of these
simulations, representing the primordial density field whose spectrum
of fluctuations is predicted by theoretical models
\cite{harrison-zeldovich}, are usually built by applying an
appropriate stochastic displacement field to extremely ordered {\em
pre-initial} configurations of the point-particle distribution
\cite{pre-initial,thierry1} (i.e., either a lattice or a particular ``glassy''
configuration).  However in literature the effect of the displacement
field is described only in approximate ways neglecting the contribution
of the internal correlations of the particle system before the
application of the displacement field \cite{BS95} and/or assuming
sufficiently small displacements \cite{zeldovich-app} using the
so-called {\em Zeldovich approximation}. In the present paper we give
the exact description of these effects at every spatial scale and for
any spatially stationary stochastic displacement field. In the
cosmological context these results can be useful for example to
understand better the role of the small spatial scales on the dynamics
of the structure formation \cite{shuf-lat,lebo,thierry2}.

Another important application deals with the problem of {\em
realizability} of point processes with an arbitrary a-priori given
two-point correlation function satisfying the hypothesis of the
Wiener-Khinchin theorem \cite{gnedenko,torquato-book}. This problem is
of great importance for instance to study the permitted disordered
configurations of hard spheres systems \cite{torquato-real}. While for
{\em continuous} stationary stochastic processes (e.g., Gaussian
processes) the hypotheses of the Wiener-Khinchin theorem give the
necessary and sufficient conditions for the realizability of the
process, this is not true for point processes. In this case the same
hypotheses, adapted to point processes, provide only necessary
conditions for the realizability of the process. Therefore finding
some limitations in the realizability of point processes can be
extremely useful for this problem. In this context we will show that
perturbing a regular lattice (which can be considered to be the most
regular and uniform point process) with any stochastic displacement
field with a continuous spectrum, it is not possible to generate a
point-particle distribution with an arbitrarily small degree of
disorder with respect to the initial lattice. In fact, we will see
that a kind of lower limit appears for the degree of disorder injected
into an initial regular lattice by any displacement field. This lower
limit is measured by a maximal finite value of the exponent of the
power spectrum of the lattice perturbed by the displacement field at
small wave numbers.

Other possible applications come from the study of the void
distribution and Voronoi tessellation in the superhomogeneous class of
point processes \cite{io-torquato}. 

The paper and the presentation of the results are organized as
follows: In Sec.~\ref{basic} basic statistical properties of point
processes are briefly presented. This includes a classification of all
spatially stationary point processes in three classes (i.e., {\em
essentially Poisson, superhomogeneous}, and {\em critical}) in terms
the asymptotic scaling behavior of the number fluctuations.

In Sec.~\ref{naive} we introduce an approximate argument, often used in
many physical applications, about the effect of the displacement field 
on the two-point correlation properties of a point process.
In this presentation we make clear that this approximation is valid
in the limits of small displacements and large spatial scales.

The rigorous treatment of the problem is introduced in
Sec.~\ref{definition}.

\section{Basic definitions}
\label{basic}

First of all let us recall some useful definitions about stochastic
mass density fields.  Given a generic (discrete or continuous)
stochastic mass density field $\rho(\bx)$ with spatially stationary
statistical properties in a $d-$dimensional Euclidean space its
average value is defined by 
\[
\left<\rho(\bx)\right>= \rho_0\,,
\]
where the symbol $\left<...\right>$ indicates the ensemble average
\footnote{In the case of ergodicity it can be taken also to be a
volume average in the infinite volume limit.}. We will limit
our analysis to the case of stochastic mass fields which can be
considered spatially uniform (elsewhere homogeneous) at sufficiently
large scale. This implies that $\rho_0>0$, excluding in this way the
case of fractal mass distributions for which $\rho_0=0$ asymptotically
but with large fluctuations in the conditional density at all scales
\cite{libro}.  

The main correlation properties of the density field
are given by the covariance function (CF), also called {\em connected}
or {\em reduced} two-point correlation function: 
\[
C(\bx-\by)=\left<\rho(\bx)\rho(\by)\right>-\rho_0^2\,.
\]
Another important quantity to characterize the internal two-point
correlation properties of a stochastic field is the so-called power
spectrum (PS) $S(\bk)$ (also called {\em structure factor}).  It
is defined by
\be
S(\bk)=\lim_{L\rightarrow+\infty}\frac{\left<|
\delta_\rho(\bk;L)|^2\right>}{L^d}\,,
\label{eq3}
\ee
where
\be
\delta_\rho(\bk;L)=\int\!\!\int_{-L/2}^{L/2}d^dx \;
[\rho(\bx)-\rho_0] e^{-i\bk\cdot\bx}\,.
\label{eq4}
\ee
Clearly in the limit $L\rightarrow+\infty$ Eq.~\ref{eq4} becomes the
Fourier transform of the {\em density contrast} $\rho(\bx)-\rho_0$.
The quantity $S(\bk)$ measures the net weight of each $k-$mode
to the determination of the stochastic process without taking into 
account the phase contribution.
  
Due to the spatial stationarity of the stochastic field,
it is simple to show that 
$S(\bk)$ is equal to the Fourier transform of $C(\bx)$
\cite{gardiner}:
\be
S(\bk)=\int d^dx\; C(\bx)e^{-i\bk\cdot\bx}
\equiv FT[C(\bx)]\,,
\label{eq5}
\ee
implying in turn
\[
C(\bx)=\frac{1}{(2\pi)^d}\int d^dk\; S(\bk)e^{i\bk\cdot\bx}
\equiv FT^{-1}[S(\bk)]\,.
\]
Note that the condition that $C(\bx)$ vanishes for 
$|\bx|\rightarrow +\infty$ implies that $k^d S(\bk)\rightarrow 0$
for $|\bk|\rightarrow 0$.
If the particle distribution is 
also statistically isotropic $C(\bx)$ depends only on $x=|\bx|$
and $S(\bk)$ on $k=|\bk|$.

We devote the rest of the paper to the so-called spatially stationary
{\em point processes} (SPP), i.e., stochastic mass fields consisting of
point-particles of unitary mass. For this class of systems the
microscopic {\em mass} density $\rho(\bx)$ coincides with the
microscopic {\em number} density $n(\bx)$ which can be written as
\be
n(\bx)=\sum_i \delta(\bx-\bx_i)\,,
\label{eq6}
\ee
where $\bx_i$ is the spatial position of the $i^{th}$ particle of the
system, $\delta(\bx)$ is the usual $d-$dimensional Dirac delta
function, and the sum is extended to all the particles of the system.
As aforementioned, we limit the discussion to SPP characterized by a
well defined average number density $n_0>0$ (i.e., excluding
fractal-like particle distributions).  Due to Eq.~\ref{eq6} and to
the fact that $n_0>0$, it is simple to find that, for a SPP, the
covariance $C({\bf x})=\left<n(\bx_0+\bx)n(\bx_0)\right>-n_0^2$ has a
{\em diagonal} singular Dirac delta-like contribution at $x=0$. That
is, it can be written as
\be
C(\bx)=n_0\delta(\bx)+n_0^2h(\bx)\,,
\label{eq7}
\ee where $n_0^2h(\bx)$ is the off-diagonal part measuring the spatial
correlation between number fluctuations in different spatial points,
i.e., for $x>0$. In general, for truly stochastic point processes,
this is a rather {\em smooth} function of $\bx$ and goes to zero for
$x\rightarrow +\infty$, but in some cases (see below the examples of
the ``shuffled'' lattices) it can present also singularities and Dirac
delta-like spikes.  Note that the spatially stationarity and the
exclusion of a fractal-like behavior imply that there is a finite
length scale $\lambda_0>0$ beyond which fluctuations of the mass
contained in volume of such size become ``small'' with respect to the
average value of the mass itself. Well beyond this distance the mass
(i.e., number) distribution can be considered with good approximation
spatially uniform or homogeneous and for this reason it is called {\em
uniformity} or {\em homogeneity} scale \cite{libro}.  It is simple to
show that in the case in which $h(\bx)$ is sufficiently regular, this
scale can be defined as the distance $\lambda_0$ such that $|h(\bx)|<
1$ for $x>\lambda_0$.  From Eqs.~\ref{eq5} and \ref{eq7} we can write
the PS of a SPP as:
\[
S(\bk)=n_0+n_0^2\hat h(\bk)\,,
\]
where $\hat h(\bk)=FT[h(\bx)]$ which in general is a regular function
decreasing to zero at large $k$. For point processes $S(\bk)$ in
statistics is also called {\em Bartlett's spectrum} \cite{bartlett}.

Since it will be useful to develop the arguments of the following
sections, we now introduce a brief classification \cite{shuf-lat} of
all the spatially stationary stochastic mass fields $\rho(\bx)$ (point
process or continuous stochastic field) with well defined $\rho_0>0$
in terms of their large scale correlations and fluctuations:\\ (1) If
$S(\bk=0)=c>0$ then $C(\bx)$ decreases to zero at large $x$ faster
than $x^{-d}$ and has a positive integral over all space (equal to
$c$), i.e., two-point correlations are short range and mainly
positive. Moreover, calling $M(R)=\int_{{\cal S}(R)}d^dx\; \rho({\bf
x})$ the mass in a given sphere ${\cal S}(R)$ of radius $R$, the
average quadratic fluctuation of this quantity behaves as
$\left<\Delta M^2(R)\right>\sim R^d$. For this reason this class of
systems can be called {\em substantially Poissonian}, as on
sufficiently large scales the system shows basically Poissonian
fluctuations. A characteristic physical example is given by a
homogeneous gas at high temperature.\\ (2) If $S({\bf k})\sim
k^{\beta}$ at small $k$ with $-d<\beta<0$ then $C(\bx)\sim
x^{-\beta-d}$ at sufficiently large $x$. That is the system has long
range and mainly positive correlations (i.e., $\int d^dx\,C({\bf
x})=+\infty$). This implies $\left<\Delta M^2(R)\right>\sim
R^{d-\beta}$ and for this reason such systems are called {\em
superPoissonian} or {\em critical}. A physical example in this class
is given by the density field of a fluid at the critical point of the
gas-liquid second order phase transition.\\ (3) If $S(\bk)\sim
k^{\beta}$ with $\beta>0$ at small $k$ then we can say that at large
$x$ the CF $C(\bx)$ decays faster than $x^{-d}$ and that $\int
d^dx\,C(\bx)=0$. This means that two-point spatial correlations are
essentially short range.  However they are not mainly positive: The
condition $S(0)=0$ indeed implies a precise balance between positive
and negative two-point correlations.  More precisely the relation
$S(0)=0$ can be seen as a condition of geometrical {\em order} in the
spatial organization of the stochastic mass fluctuations. The higher
is $\beta$, the higher the large scale degree of order.  As a matter
of fact, as shown below, in the case of a regular and periodic
lattice of particles, which is the most ordered particle distribution,
one can say that $S(\bk)\sim k^\beta$ with $\beta\rightarrow +\infty$
for $k\rightarrow 0$. For all these reasons for this last class of
mass fields the name {\em superhomogeneous} has been proposed
\cite{shuf-lat,lebo,libro}  (elsewhere {\em hyperuniform}
\cite{torquato-real}).

\section{An approximate argument}
\label{naive}

Before entering the detailed and rigorous discussion, we give an
argument usually used to roughly describe the effect of a displacement
field on a ``sufficiently uniform'' mass distribution (for another
approximate result to this problem going further than the present
approximation see \cite{BS95}).  This argument is based on the fact
that, if the applied displacements are sufficiently small, the mass is
conserved ``locally''. Hence a form of {\em continuity equation} has
to be satisfied. Let us call $\rho_{in}(\bx)$ the initial microscopic
density field, and $\rho(\bx)$ the same quantity after the application
of the displacement field $\bu(\bx)$. By considering the displacements
``small'' enough, we can write the one-step continuity equation:
\be 
\rho(\bx)-\rho_{in}(\bx)+{\bf
\nabla}\cdot\left[\rho_{in}(\bx) \bu(\bx)\right]\simeq 0\,,
\label{int-disp1}
\ee
where $\bu(\bx)$ is the displacement performed at point $\bx$ in the given
temporal step. The equality is rigorously satisfied only in the case 
of infinitesimal displacements.
Let $\rho_0>0$ be 
\[\rho_0=\left<\rho_{in}(\bx)\right>\,\]
It is simple to verify that this average value is not modified 
by the action of the displacement field. If $\rho_{in}(\bx)$ is
``sufficiently uniform'' with respect to $\rho(\bx)$, we can
approximate it in Eq.~\ref{int-disp1} with a continuous and uniform 
density field $\rho_{in}(\bx)=\rho_0$, so that Eq.~\ref{int-disp1} 
can be rewritten as
\be
\rho(\bx)-\rho_0+\rho_0{\bf \nabla}\cdot\bu(\bx)\simeq 0\,.
\label{int-disp2}
\ee
By taking the Fourier integral in a cubic volume of size $L$ of 
Eq.~\ref{int-disp2}, and using Eq.~\ref{eq4}, we have:
\be
|\delta_\rho(\bk;L)|^2\simeq 
\rho_0^2|\bk\cdot\bv(\bk;L)|^2\,,
\label{int-disp3}
\ee
where 
\[\bv(\bk;L)=\int\!\!\int_{-L/2}^{L/2} d^dx\, 
e^{-i\bk\cdot\bx}\bu(\bx)\,.\] 
From Eqs.~\ref{int-disp3} and \ref{eq3}, we can say that the PS of the
final mass distribution is roughly proportional to $k^2$ times the PS
of the displacement field (actually the PS of the vector field
$\bu(\bx)$, as its CF, is a $d\times d$ matrix and the form of such
relation can be more complicated).

This simple result is based on two assumptions: the former consists
in approximating the initial microscopic density with a completely
uniform continuous mass field.  As a consequence we expect that in the
exact equations describing the effect of the displacement field on a
spatial distribution of identical particles, there will be terms
related to the inhomogeneities (e.g., ``granularity'') of the initial
particle density. The latter is the fact that, behind
Eq.~\ref{int-disp1}, there is the assumption of ``sufficiently small''
displacements. Then we expect that, when this assumption is not valid,
the relation Eq.~\ref{int-disp3} will change drastically.

\section{Definition of the problem}
\label{definition}

We start by considering a SPP, as defined above, with
microscopic density $n_{in}(\bx)$ given by Eq.~\ref{eq6} and with $n_0>0$.

Let us now suppose of introducing a stochastic
displacement field $\bu(\bx)$ displacing each particle from its
initial position.  In general this displacement process changes the PS
of the initial point distribution from $S_{in}(\bk)$ to a new $S({\bf
k})$ (or equivalently the initial CF $C_{in}(\bx)$ to a new $C(\bx)$).
If $\bu_i$ is the displacement applied to the particle $i$, the
position of this particle passes from $\bx_i$ to $\bx_i+\bu_i$ (see
Fig.\ref{config}). Therefore the final particle density field can be
written as
\be
n(\bx)=\sum_i \delta(\bx-\bx_i-\bu_i)\,.
\label{disp2}
\ee

\bef
\includegraphics[height=6.5cm,width=8.5cm,angle=0]{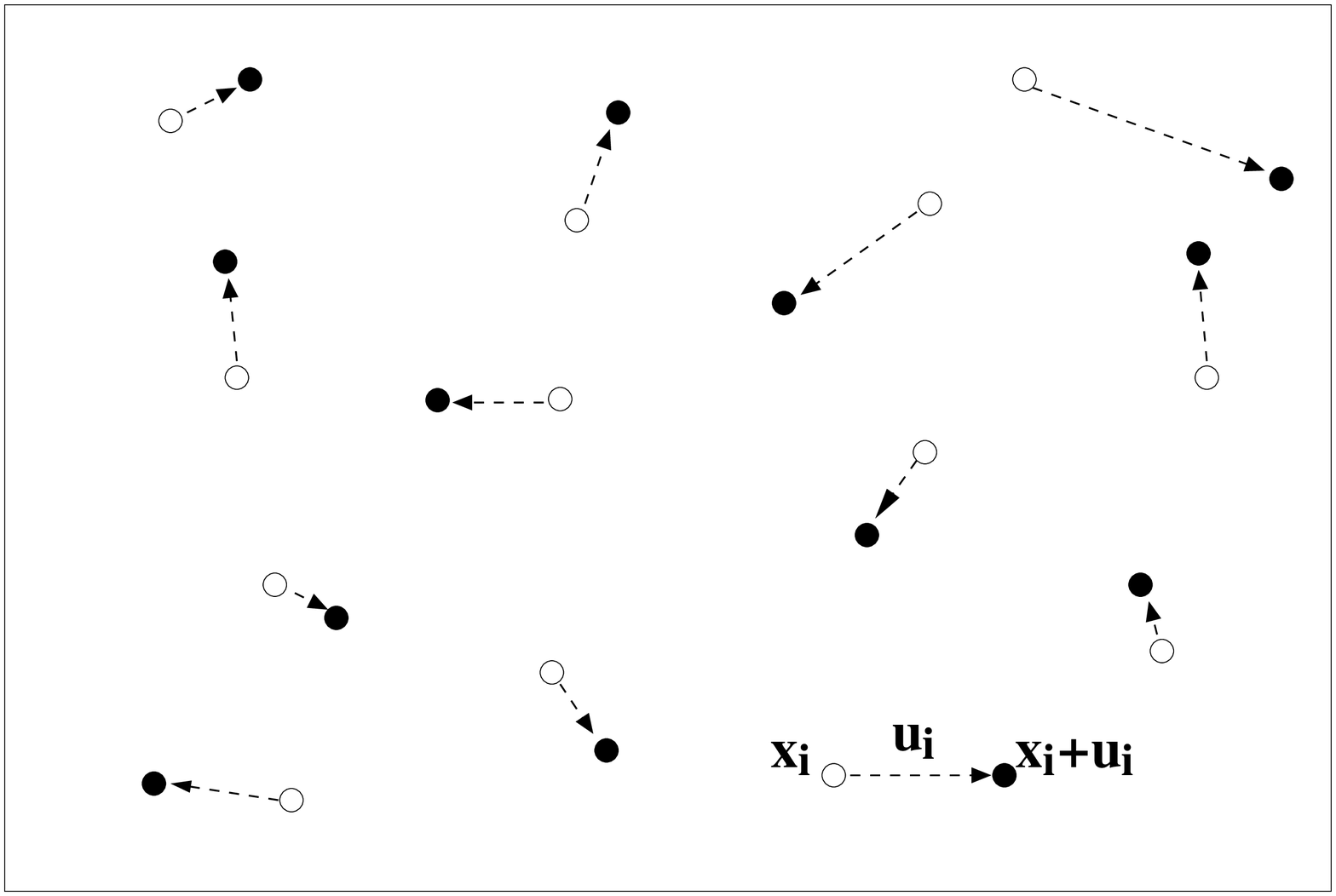}
\caption{The figure presents a pictorial view of the effect of a
stochastic displacement field to a spatial particle distribution in
$2d$.  The particles pass, through the displacements (dashed arrows),
from the old positions (black circles) to the new ones (empty circles).
\label{config}}
\eef

A stochastic displacement field $\bu(\bx)$ can be seen as a {\em
continuous} stochastic vector field. We assume that this field is
spatially stationary in the statistical sense, i.e., it is
characterized by the invariance of the statistical properties for any
spatial translation. We can think to ``attach'' a
displacement vector $\bu(\bx)$ to each spatial point $\bx$, even
though it acts on the mass density only if $\bx$ is occupied by a
particle.  In what follows, we assume that the displacement field is
statistically independent of the realization $n_{in}(\bx)$ of the
initial particle density, i.e., the probability of having a given
realization $\bu(\bx)$ of the displacement field is independent on the
realization of the initial particle distribution.

Let us consider a function $A$ only of the displacements
$\{\bu_1,...,\bu_N\}$ applied respectively to a set of spatial points
$\{\bx_1,...,\bx_N\}$. The average of this quantity over all the
realizations of the displacement field $\bu(\bx)$ is defined by
\be 
\overline{A}=\int\!..\!\int
\left[\prod_{j=1}^N d^du_j\right] f_N(\bu_1,...,\bu_N) 
A(\bu_1,...,\bu_N)\,,
\label{disp3b}
\ee 
where $f_N$ is the joint probability density function (PDF) of the
displacements $\{\bu_1,...,\bu_N\}$ applied respectively to the set of
points $\{\bx_1,...,\bx_N\}$. In general, $f_N$ depends parametrically
on the points positions $\bx_i$.  In the case of a statistically
stationary displacement fields, $f_N$ depends parametrically only on
the separation vectors between all the couples of the points of the
set $\{\bx_1,...\bx_N\}$.  A particular and very important case is
when the set of points coincides with the positions occupied by all
the particles (in which case we call the joint displacement PDF simply
${\cal P}(\{\bu_i\})$) of the initial SPP or by the particles of one
of its subsets.  Note, however, that in our hypothesis the form of
this PDF does not depend on the fact that these points are actually
occupied by particles.

Finally, if we have a function of the final (i.e., after the
introduction of the displacements) microscopic density $n(\bx)$, the
ensemble average over all the possible final configurations of the
particle distribution is given by averaging over all the possible
displacements as in Eq.~\ref{disp3b}, fixing the initial particle
density $n_{in}(\bx)$, and then over all the possible initial particle
configurations $\left<...\right>$.  This is due to the fact that the
{\em ensemble} of the all possible final particle configurations is
found by considering all the possible initial configurations, and for
each of these all the final configurations obtained by applying the
{\em ensemble} of the displacement fields.  However, if, as we suppose
here, the displacement field is statistically independent on the
initial particle distribution, the order of these two
averages is arbitrary. For instance, in this case, the CF of the 
``displaced'' particle distribution is expressed by
\be C(\bx)=\left<\overline{n(\bx_0+\bx)n(\bx_0)}
\right>-\left<\overline{n(\bx_0)}\right>^2\,,
\label{disp3c}
\ee
with an arbitrary order of the two averages $\overline{(...)}$ and
$\left<...\right>$.

\section{Exact results for one and two-point statistical
properties of the particle distribution}
\label{exact1}

The aim of this section is to relate the one and two-point
correlation properties of the ``displaced'' particle distribution to its 
initial ones and to those of the applied displacement field by finding exact
relations going beyond the approximation given by Eq.~\ref{disp3c}.
As aforementioned, the discussion will be limited to the case of
spatially stationary stochastic displacement fields and initial particle
distributions. In this way also the final particle distribution will be
spatially stationary.

We will start by evaluating, through Eq.~\ref{disp2}, the average mass
density $\left<\overline{n(\bx)}\right>$. The next step will
consist in finding the transformation equation for the PS $S(\bk)$
(or equivalently the CF $C(\bx)$).

Since the displacement process does not create or destroy any
particle and is statistically stationary, 
the average mass density stays equal to the initial one
$n_0$:
\[
\left<\overline{n(\bx)}\right>=n_0\,.
\]
This can be also proved by direct calculation using
$\left<n_{in}(\bx)\right>=n_0$. First of all we note that
Eq.~\ref{disp2} is a sum of single displacement terms. Therefore, in
order to evaluate the displacement average $\overline{n(\bx)}$, we
need only to know the one displacement PDF $f_1(\bu)$ and not the
complete joint PDF ${\cal P}(\{\bu_i\})$ for all the system
particles. In our hypothesis of spatial stationarity $f_1(\bu)$ does
not depend on the point of application of the displacement, and we
recall it $p(\bu)$ for simplicity ($p(\bu)$ is obtained from ${\cal
P}(\{\bu_i\})$ by integrating out all the displacements with the exception
of one).  We can then write
\[
\overline{n(\bx)}=\sum_i \int d^du_i\,p(\bu_i)
\delta(\bx-\bx_i-\bu_i)=\sum_i p(\bx-\bx_i)\,.
\]
By taking the average $\left<...\right>$ over the initial particle
configurations, we finally have
\bea
&&\left<\overline{n(\bx)}\right>=
\left<\int d^dy\,p(\by)
\sum_i \delta(\by-\bx+\bx_i)\right>=\nonumber\\
&&n_0
\int d^dy\,p(\by)=n_0\,,\nonumber
\eea
where we have used the statistical spatial stationarity of 
$n_{in}(\bx)$ (i.e., $\left<n_{in}(\bx)\right>=
\left<n_{in}(\by-\bx)\right>=n_0$) and the normalization
condition of the one-displacement PDF $p(\bu)$.

We can now face the problem of calculating the new CF $C(\bx)$ and
the new PS $S(\bk)$.  The key point is to evaluate the average
$\left<\overline{n(\bx)n(\by)}\right>$. Since the product
\[n(\bx)n(\by)=\sum_{i,j}\delta(\bx-\bx_i-\bu_i)
\delta(\by-\bx_j-\bu_j)\,,\] 
is a sum of terms containing either one (for $i=j$) or two (for $i\ne
j$) different displacements, we do not need to know the complete joint
PDF ${\cal P}(\{\bu_i\})$,
but only the joint two-displacement PDF $f_2(\bu,\bv)$ which is
obtained from ${\cal P}(\{{\bf u}_i\})$ by integrating out all the
displacements but two.  In general
$f_2(\bu,\bv)$ will depend parametrically on the coordinates of
the two points of application of the displacements.  Assuming the hypothesis
of a spatially stationary displacement field, $f_2(\bu,\bv)$ depends 
parametrically only on the separation vector $\bx$ between
these two points. For this reason we recall it $f_2(\bu,{\bf
v})\equiv f(\bu,\bv;\bx)$ putting in explicit evidence this
dependence.  Note that the function $f(\bu,\bv;\bx)$ carries much more
information than the simple knowledge of the average displacement
$\overline{\bu}={\bf U}$, and the two-displacement correlation matrix of
elements
\be
G_{\mu \nu}(\bx-\by)=\overline{\left(u^{(\mu)}(\bx)-
U^{(\mu)}\right)\left(u^{(\nu)}(\by)-U^{(\nu)}\right)}
\label{disp-corr}
\ee
with $\mu, \nu=1,...,d$, where $u^{(\mu)}$ is the $\mu^{th}$ component
of the displacement vector $\bu$.  In fact $G_{\mu \nu}({\bf
x}-\by)$ is only the average value of $\left(u^{(\mu)}(\bx)-
U^{(\mu)}\right)\left(u^{(\nu)}(\by)-U^{(\nu)}\right)$ calculated
with the PDF $f(\bu,\bv;\bx)$ itself, while the knowledge
of $f(\bu,\bv;\bx)$ include all information about all the higher moments
of two-displacements.

The joint two-displacement PDF $f(\bu,\bv;\bx)$ satisfies
the following limit conditions in $\bx$:
\bea
&&f(\bu,\bv;0)=\delta(\bu-\bv)p(\bu) 
\label{disp4a}\\
&&\lim_{x\rightarrow\infty} f(\bu,\bv;\bx)=p(\bu)p(\bv)
\label{disp4b}
\eea 
The former equation is trivial, while the latter establishes simply
that the correlation between two different displacements must go to
zero if the distance between the two points of application goes to
infinity. 

First of all let us evaluate the average of $n(\bx)n(\by)$ over the
displacements. By direct integration one obtains
\bea
\label{disp5}
&&\overline{n(\bx)n(\by)}=\\
&&\sum_{i,j}\!\int\!\!\int d^du_i d^du_j
[f(\bu_i,\bu_j;\bx_{ij})\delta(\bx-\bx_i-\bu_i)
\times\nonumber\\ 
&&\delta(\by-\bx_j-\bu_j)]=
\sum_{i,j} 
f(\bx-\bx_i,\by-\bx_j;\bx_{ij})\,,\nonumber
\eea 
where $\bx_{ij}=\bx_i-\bx_j$.  Note that the first limit condition in
Eq.~\ref{disp4a} permits to perform the average without separating the
diagonal contribution $i=j$ from the off-diagonal part $i\ne j$ of the
double sum in Eq.~\ref{disp5} by averaging separately the former using
the one-displacement PDF $p(\bu_i)$ and the latter through the
two-displacement PDF $f(\bu_i,\bu_j;\bx_{ij})$ with $i\ne j$.

The next step is to evaluate the average $\left<...\right>$ on the
ensemble of initial particle configurations. For this purpose note that
the ensemble average of a function of the initial
configuration of the form
$\sum_{i,j}\psi(\bx_i,\bx_j)$, where $\psi(\bx,\by)$ is a
generic two-point function, can be written as
\bea
\label{disp6}
&&\left<\sum_{i,j}\psi(\bx_i,\bx_j)\right>
\equiv\\
&&\left<\int\int  d^dx d^dy\;\psi(\bx,\by) 
\sum_{i,j}\delta(\bx-\bx_i)\delta(\by-\bx_j)\right>=
\nonumber\\
&&\int\int  d^dx d^dy\;\psi(\bx,\by)
\left<\sum_{i,j}\delta(\bx-\bx_i)\delta(\by-\bx_j)\right>=
\nonumber\\
&&\int\int d^dx d^dy\; \left<n_{in}(\bx)n_{in}(\by)\right> 
\psi(\bx,\by) \,,\nonumber
\eea
where, by definition, we have
\be
\left<n_{in}(\bx)n_{in}(\by)\right>=n_0^2+
C_{in}(\bx-\by)\,.
\label{disp7}
\ee
Note that the diagonal part $n_0\delta(\bx)$ of the connected
two-point correlation function $C_{in}(\bx)$ takes correctly
into account the diagonal term $i=j$ of the sum of Eq.~\ref{disp6}.

By applying Eqs.~\ref{disp6} and \ref{disp7} to Eq.~\ref{disp5},
it is possible to write:
\bea
&&\left<\overline{n(\bx)n(\by)}\right>=\int\!\!\int 
d^dx' d^dy'\left[n_0^2+C_{in}(\bx'-\by')\right]\nonumber\\
&&\times f(\bx-\bx',\by-\by';\bx'-\by')
\label{disp7b}
\eea 
It can be convenient to rewrite Eq.~\ref{disp7b} by separating the two
terms coming respectively from the diagonal and the off-diagonal parts
of $C_{in}(\bx)$, that is by writing, as in Eq.~\ref{eq7},
$C_{in}(\bx)=n_0\delta(\bx)+n_0^2 h_{in}(\bx)$:
\bea
\label{disp7bb}
&&\left<\overline{n(\bx)n(\by)}\right>=n_0\delta(\bx-\by)\\
&&+n_0^2\int\!\!\int d^dx' d^dy'
\left[1+h_{in}(\bx'-\by')\right]\nonumber\\
&&\times f(\bx-\bx',\by-\by';\bx'-\by')\,.\nonumber
\eea
We are now able to write the new CF $C(\bx)$ of the final
particle distribution which is defined, as usual, by 
\be
C(\bx)=\left<\overline{n(\bx_0+\bx)
n(\bx_0)}\right>-n_0^2\,.
\label{disp7bc}
\ee 
Note that from Eq.~\ref{disp7bb} the diagonal part of $C(\bx)$ remains
equal to that of $C_{in}(\bx)$ (i.e., $n_0\delta(\bx)$) as expected.

In order to write the transformation equation for the PS, which is the
most useful in many applications \cite{pre-initial}, we start from the
simplest case of uncorrelated displacements; then we will come back to
the general case for general considerations and some paradigmatic
examples.

\section{Independent displacements}
\label{indip}

We now consider the case in which the displacement applied to a given
spatial point is statistically independent of the displacement applied
to any other point.  Therefore the statistics of the stochastic
displacement field is completely determined by the knowledge of the
reduced one-displacement PDF $p(\bu)$, and the joint PDF of $n$
displacements $\bu_1,\bu_2, ..., \bu_n$ in $n$ {\em different} points
of the space factorizes as follows:
\[
f_n(\bu_1,\bu_2, ..., \bu_n)=\prod_{i=1}^n p(\bu_i)\,.
\]
In particular for the two-displacement PDF, we can write
\be
f(\bu,\bv;\bx)=\left\{
\begin{array}{ll}
\delta(\bu-\bv)p(\bu)&\mbox{for}\;x=0\\
\\
p(\bu)p(\bv)&\mbox{for}\;x\ne0
\end{array}
\right.
\label{disp9}
\ee
Note that the lack of displacement-displacement correlations implies
a discontinuity of $f(\bu,\bv;\bx)$ at $x=0$. As
shown below this does not happen for truly continuous correlated
stochastic displacement fields (i.e., belonging to the class of
continuous stationary stochastic processes \cite{gnedenko}).  We can
now apply Eq.~\ref{disp9} to Eq.~\ref{disp7bb} in order to find the
two-point correlation function of the final system:
\bea
\label{disp7c}
&&\left<\overline{n(\bx)n(\by)}\right>
=n_0^2+n_0\delta(\bx-\by)\\ &&+n_0^2\int\int d^dx' d^dy'\;
p(\bx-\bx')h_{in}(\bx'-\by')p(\by-\by')\,.  \nonumber 
\eea 
Since in this case $f(\bu,\bv;\bx)$ is discontinuous at $x=0$, it has
been important to separate the contributions of the diagonal and the
off-diagonal part of $C_{in}(\bx)$ in Eq.~\ref{disp7bb}. In fact, in
the hypothesis of uncorrelated displacements, any element of the
connected two-displacement correlation matrix $G_{\mu\nu}(\bx)$, given
in Eq.~\ref{disp-corr}, vanishes for any $x\ne 0$ while
$G_{\mu\nu}({\bf 0})=\delta_{\mu\nu}g_\mu^2$ where
$g_\mu^2>0$ is the single displacement variance.
That is, $G_{\mu\nu}(\bx)$ is discontinuous at $x=0$.  As
aforementioned, this is a very particular case, as in truly
correlated continuous stationary stochastic processes it is continuous
everywhere \cite{gnedenko}.

At this point, by remembering that the PS
$S(\bk)=FT[C(\bx)]$, with $C(\bx)$ given
by Eq.~\ref{disp7bc}, we can Fourier transform Eq.~\ref{disp7c} to 
obtain the following local relation in $k$ space:
\be
S(\bk)=n_0(1-|\hat{p}(\bk)|^2)+
|\hat{p}(\bk)|^2 S_{in}(\bk)\,,
\label{disp10}
\ee
where $\hat{p}(\bk)$ is the characteristic function of the
one-displacement PDF
\[
\hat{p}(\bk)=FT[p(\bu)]\,,
\] 
and where we have used
$n_0^2FT[h_{in}(\bx)]=S_{in}(\bk)-n_0$.  
Note that by definition  $\hat{p}(0)=1$.

Equation \ref{disp10} gives the relation between the PSs of the
point-particle configurations before and after the application of the
uncorrelated displacements field.  First of all let us analyze the
notable case in which the initial point-particle distribution is the
statistically stationary and isotropic Poisson one, i.e., that case in
which there is no correlation between the initial positions of
different particles.  It is simple to show
\cite{shuf-lat,libro} that the initial density CF is simply
$C_{in}(\bx)=n_0\delta(\bx)$ (i.e., it has only the diagonal
part).  This means that $S_{in}(\bk)=n_0$ which, in view of
Eq.~\ref{disp10}, implies $S(\bk)=n_0$ too, regardless to the
form of $p(\bu)$. That is, the particle distribution remains
Poissonian after the application of any random and uncorrelated
displacement field.  This can be considered as a formulation of the
so-called {\em theorem of Bartlett} \cite{bart-th1} (for the behavior
of a Poisson point process under a {\em deterministic} displacement
field see \cite{bart-th2}).
This property is quite easy to understand: the displacement field has no
spatial correlation, hence it tends to randomize the particle
distribution, but the Poisson particle distribution is already the
``most random possible'' SPP. This is further clarified by observing
that uncorrelated displacements cannot increase the spatial
correlations in the particle distribution.  We have just shown that
the stationary Poisson point process of average density $n_0>0$ is the
``fixed point'' of the transformation given by
Eq.~\ref{disp10}.  We show now that this
fixed point (i.e., the stationary Poisson point process) is also {\em
attractive}.  That is, we show that, starting from an arbitrary
stationary point process with density $n_0$ and initial PS
$S_{in}(\bk)$, by applying successive stochastic uncorrelated
displacements to all the particles taken from the same PDF $p({\bf
u})$ with no correlation at each step, the PS flows toward
$S_{\infty}(\bk)=n_0$, i.e., the Poisson one.
It is simple to show that after $m$ consecutive applications of
such displacement field the PS $S_m(\bk)$ satisfies the relation:
\[
S_m(\bk)=n_0(1-|\hat p(\bk)|^{2m})+|\hat p(\bk)|^{2m}S_{in}(\bk)\,.
\]
The previous equation is simply obtained by the fact that the
characteristic function of the sum of $m$ independent random vectors
extracted from the same PDF $p(\bu)$ is simply $[\hat p(\bk)]^m$.
Since $p(\bu)$ is a PDF, a part from particular cases \footnote{E.g.,
the case in which $p(\bu)=\delta(\bu-\bu_0)/2+\delta(\bu+\bu_0)/2$ for
which $\hat p(\bk)=\cos(\bk\cdot\bu_0)$ and then $|\hat p(\bk)|=1$ for
all $\bk$ such that $\bk\cdot\bu_0=n\pi$ with $n$ any integer.}, in
general $|\hat p(\bk)|<1$ for any $k>0$.  This implies that $S_m(\bk)$
converges exponentially fast to $n_0$ for each $\bk$, i.e., to the PS
of the Poisson point process
\footnote{As $|\hat p({\bf 0})|=1$, for $k=0$ instead
we find singularly $S_m({\bf 0})=S_{in}({\bf 0})$ at any $m$.  This
creates in general an asymptotic isolated discontinuity at $k=0$.
However this does not matter for the real space correlation properties
as the Fourier transform of the PS is insensible to that.}.

One has also to notice that the right hand side of Eq.~\ref{disp10}
is the sum of two terms: the former is proportional to the average 
density of particles $n_0$, that we can call {\em
granularity} term, and independent of the initial PS of the particle
distribution, while the latter depends on $n_0$, only through
$S_{in}(\bk)$ which satisfies the condition
$S_{in}(\bk\rightarrow \infty)=n_0$ because of the diagonal
term of $C_{in}(\bx)$ and of the fact that the off-diagonal part
must be integrable at small $x$.

If the initial point process is statistically isotropic as well as
stationary, then $C_{in}(\bx)$ is a function only of $x=|\bx|$ and
$S_{in}(\bk)$ only of $k=|\bk|$.  Furthermore, if also the
displacement field is statistically isotropic, then
$\overline{u^{\mu}}=0$ for each $\mu=1,...,d$, and $p(\bu)$ depends
only on $u=|\bu|$.  This implies that also $S(\bk)$ depends only on
$k$ (and $C(\bx)$ on $x$).

\subsection{Small $k$ expansion and large scale behavior - I}
\label{as1a}

In many applications, such as cosmological studies \cite{n-body1}, it is
particularly important to analyze the behavior of $S(\bk)$ at small
$k$, that is at large spatial scales.  For this reason we now study
the asymptotic behavior of Eq.~\ref{disp10} for $k\rightarrow 0$.  We
limit the discussion to the case in which both the point process
generating the initial particle distribution and the stochastic
displacement field are statistically isotropic.  As seen above,
this hypothesis implies $S_{in}(\bk)=S_{in}(k)$, $p(\bu)=p(u)$, and
$S(\bk)=S(k)$.

The first step consists in studying the small $k$ behavior of the
characteristic function $\hat p(\bk)$. By definition we have
\[
\hat p(\bk)=\int d^du\;e^{-i\bk\cdot\bu}p(\bu)\,,
\]
and then $\hat p(0)=1$.  As $p(\bu)=p(u)$,
then $\hat p(\bk)=\hat p(k)$.  Let us suppose that at sufficiently 
large $u$ we have $p(u)= Au^{-\alpha}+o(u^{-\alpha})$,
where $\alpha>d$ as $p(u)$ must be by definition integrable over
all the space ($\alpha\rightarrow+\infty$ includes exponential-like or
faster decay at large $u$).  Using this property and the definition of
$\hat p(k)$, it is simple to show
that to the lowest order in $k$ larger
than zero, one has
\be
\hat p(k)= 1 - Bk^{\beta}\;\mbox{with}\,\left\{
\begin{array}{ll} 
\beta=\alpha-d &\mbox{if}\;0<\alpha-d\le 2\\
\\
\beta=2&\mbox{if}\;\alpha>d+2\\
\end{array}
\right.
\label{as1-2}
\ee
where $B>0$.  
Moreover if $\alpha>d+2$ then $B=\frac{\overline{u^2}}{2d}$.
Instead, if $d<\alpha\le d+2$, then $\overline{u^2}$ diverges and
\be
B=A\int d^dx\, x^{-\alpha}\left(1-e^{-ix\cos\theta}\right)\,, 
\label{as1-3}
\ee
where $\theta$ is the angle between $\bx$ and any one of the 
coordinate axes.
Note that in any case $0<\beta\le 2$.  This implies that for 
$k\ll B^{1/\beta}$
\[1-|\hat p(k)|^2\simeq 2B\,k^\beta\,.\]
On the other hand, as seen in Sec.~\ref{basic}, in order to have the
initial SPP well defined, it is necessary that $k^d
S_{in}(k)\rightarrow 0$ for $k\rightarrow 0$, that is practically
$S_{in}(k)\sim k^{\gamma}$ at small $k$ with $\gamma>-d$.
 
We can draw the following
conclusions for the small $k$ behavior of $S(k)$ in Eq.~\ref{disp10}:
\begin{enumerate}
\item If $-d<\gamma<\beta$ (with as seen above $0<\beta\le 2$),
$S(k)\sim k^{\gamma}$ similarly to $S_{in}(k)$, and its small $k$
amplitude is independent of the displacement field. This means that,
in this case, finite {\em uncorrelated} displacements cannot destroy
the persistence of correlations already present in the system.  In
particular it is important to note that if $\gamma<0$ (i.e., long
range density-density correlations in the initial particle
configurations) no uncorrelated displacement field is able to affect
the large scale correlation properties of the initial system. 
\item On the contrary, if $0<\beta< \gamma$, then the small $k$
behavior is completely determined by the displacement field, resulting
in $S(k)\simeq 2Bk^\beta$. As shown in Sec.~\ref{basic}, a
point-particle distribution having $\gamma>0$ is called {\em
superhomogeneous} showing a sort of long range order, characterized by
{\em subPoissonian} mass fluctuations at large scales, which in the
present case is partially destroyed by the Poissonian noise introduced
by the displacement field.  In this respect, note that if
$\overline{u^2}=+\infty$ then $\beta<2$.  Consequently, it is much
simpler to obtain the condition $\gamma> \beta$, i.e., the
randomization of the system introduced by the uncorrelated
displacement field is more effective.
\item If $\beta=\gamma$ both the long wave length modes of the initial
particle configuration and of the displacement field
determine of the small $k$ power spectrum of the final system. In
particular the exponent is equal to the initial one while its
amplitude increases. This indicates that the initial and the final
systems have the same kind of long range order of density fluctuations
and the same mass-length scaling relation for the large scale mass
fluctuations, but with an increase of their amplitude in the second case.
\end{enumerate}

\subsection{The shuffled lattice with uncorrelated displacements}

In this subsection we present a simple, but important example of
application of uncorrelated stochastic displacement fields: the random
shuffling of a regular lattice of particles.  Its importance relies on
the fact that a {\em perturbed} lattice is often used as the initial
condition for many dynamical applications as, for instance, the already
mentioned cosmological $n-$body simulations \cite{pre-initial}, and
bio-metrical studies
\cite{renshaw}.
In this section we present the simplest example of a lattice with
stochastic displacement perturbations. In Fig.~\ref{sl-3d} the
projection on the $xy$ plane of a three dimensional shuffled lattice
is given.

\bef
\includegraphics[height=7.5cm,width=7.5cm,angle=0]{fig2.eps}
\vspace{0.6cm}
\caption{The projection on the $xy$ plane of a $3d$ {\em shuffled}
lattice of $16^3$ particles in the unitary volume is represented. In
the present case each particle is randomly displaced from its initial
lattice position inside a finite cubic box centered around the lattice
point and of side equal to one fifth of the lattice spacing. The
displacement applied to each particle is statistically independent of
the displacements applied to the other particles.
\label{sl-3d}}
\eef

It is simple to show that for a distribution of point particles
of unitary mass occupying the sites of a regular cubic lattice, the PS
is \cite{ziman} 
\be
S_{in}(\bk)=(2\pi)^d n_0^2 \sum_{\bH\ne {\bf 0}}
\delta\left(\bk-\bH\right)\,,
\label{sh-lat1}
\ee
where the sum runs over all the sites $\bH$ of the {\em reciprocal}
lattice \cite{ziman,ashcroft} with the exception of the origin ${\bf 0}$.
We remind that, if the particles occupy the sites of a cubic lattice
with lattice spacing $l$, then each component of $\bH$ is a positive
or negative integer multiple of $2\pi/l$ and the average density of
points is $n_0=l^{-d}$.

We can now apply Eq.~\ref{disp10} in order to find the final PS
$S(\bk)$ after the random shuffling (i.e., the application of the
random displacement field):
\be
S(\bk)=n_0\!\left(1-\left|\hat p(\bk)\right|^2\right)+
(2\pi)^d n_0^2\!\sum_{\bH\ne {\bf 0}}\left|\hat p(\bH)\right|^2
\!\delta\left(\bk-\bH\right)\,,
\label{sh-lat2}
\ee
\bef [tbp]
\includegraphics[height=7cm,width=8.5cm,angle=0]{fig3.eps}
\vspace{0.2cm}
\caption{Power spectrum of a $1d$ shuffled lattice with uncorrelated
displacements with finite variance obtained both from numerical
simulations and from the theoretical prediction given by
Eq.~\ref{sh-lat2} and showing perfect agreement.  In particular the
two PSs refer to a regular chain of particles and unitary lattice
spacing perturbed by the displacement field, through which each
particle is randomly displaced in a point of the segment centered
around its initial lattice position and of length $a=1/50$ (i.e.,
$p(u)=\theta(a/2-u)\theta(u+a/2)/a$). Note that at small $k$ the PS
$S(k)$ scales as $k^2$ as the displacement variance is
finite. Moreover the Bragg peaks (whose amplitude here has been
normalized for pictorial reasons) are perfectly modulated by weights
proportional to $|\hat p(k)|^2$ (which in this case is given by $\hat
p(k)=2\sin^2(ka/2)/ka$). Finally the PS at large $k$ converges
correctly to the average number density $n_0=1$.
\label{sl-k2}}
\eef 

We now stress two important aspects of Eq.~\ref{sh-lat2}:
\begin{enumerate}
\item The random shuffling in general does not erase completely the
presence of the so-called {\em Bragg peaks} (i.e., the sum of delta
functions), but only modulates their amplitude and adds a continuous
contribution typical of fully stochastic point
distributions. The complete cancellation of the Bragg peaks
contribution to $S(\bk)$ is possible only in the very particular
case in which $\hat p(\bH)=0$ for every reciprocal lattice vector
with the exception of ${\bf 0}$.
\item around $k=0$ (more precisely in the so-called {\em first
Brillouin zone} \cite{ashcroft} of the reciprocal lattice)
$S_{in}(\bk)=0$ at any order of the Taylor expansion (we can loosely say 
that $S_{in}(\bk)\sim k^{+\infty}$ around $k=0$).  Consequently,
as clear from Eq.~\ref{sh-lat2}, in this region $S(\bk)$ is determined
by only the behavior of the displacement characteristic function $\hat
p(\bk)$.  As shown above, if the displacement field is statistically
isotropic, $\hat p(\bk)\equiv\hat p(k)$.  Therefore, even though the
lattice is strictly anisotropic, the shuffled one has isotropic mass
fluctuations at large scales. This implies that on sufficiently large
scales the scaling exponent of the fluctuations of the
number of particles contained in a given volume of linear size $R$ do
not depend strongly on the shape of the volume itself.  This is not
true, instead, for a deterministic cubic lattice for
which the scaling behavior of these fluctuations changes in passing
from a spherical to a cubic volume with the same symmetry of the
lattice.
\end{enumerate}
Since in the first Brillouin zone $S(\bk)$ is completely determined by
$\hat p(\bk)$, the asymptotic behavior at small $k$ of $S(\bk)$ can be
derived by Eq.~\ref{as1-2}.  In particular if the variance
$\overline{u^2}$ of the displacement field is finite, we find
$S(\bk)\sim k^2$ for $k\rightarrow 0$, independently on the particular
form of $p(u)$ (see Fig.~\ref{sl-k2}). 
This is a case of {\em universal behavior} for all the
PDF $p(u)$ with sufficiently fast decay at large $u$. 

Instead, in the case in which $\overline{u^2}$ diverges, i.e.,
$d<\alpha\le d+2$, this universality is lost, having $S(\bk)\sim
k^\beta$ with $\beta=\alpha-d$, with a one-to-one correspondence
between the exponents $\alpha$ and $\beta$ as shown in the previous
section (see Fig.~\ref{sl-k}).  A similar case of universality is
found in random walks with independent steps \cite{random-levy}. In
fact if the variance of the steps is finite (ordinary random walks)
the average quadratic distance $\left<\Delta x^2(t)\right>$ reached by
the walker after a large number $t$ of steps satisfies the scaling
relation $\left<\Delta x^2(t)\right>\sim t$ independently of the
precise functional form of PDF of the single step. On the other hand,
if the single step variance goes to infinity (Levy flights) this is no
more true, $\left<\Delta x^2(t)\right>$ being infinite, and the PDF of
$\Delta x(t)$ at sufficiently large $t$ having a power law tail with
an exponent in a one-to-one correspondence with that characterizing
the tail of the PDF of a single step. A similar transition from a universal
scaling behavior of the fluctuations to a non-universal one has been
found also in more complex fragmentation problems \cite{k4}.
\bef [tbp]
\includegraphics[height=6.5cm,width=8cm,angle=0]
{fig4.eps}
\vspace{0.2cm}
\caption{Shuffled lattice of the same kind of that in
Fig.~\ref{sl-k2}, but now with $p(u)=(a/\pi)/(u^2+a^2)$ where
$a=1/50$, i.e., with unlimited displacements and logarithmically
diverging $\overline{u}$. Again the agreement between the numerical
simulation results and the theoretical prediction Eq.~\ref{sh-lat2} is
excellent. Note in particular that $S(k)\sim k$ at small $k$ and that
the remaining Bragg peaks, superimposed to the continuous contribution
to the PS, are well modulated by weights proportional to $|\hat
p(k)|^2$ with $\hat p(k)=\exp (-a|k|)$.  Finally the PS at large $k$
converges to average number density $n_0=1$.
\label{sl-k}}
\eef

\section{Correlated displacements}
\label{correl}
 
Let us now go back to Eq.~\ref{disp7bb}, and consider the general
case of a stationary stochastic displacement field with spatial
correlations.  In this case $f(\bu,\bv;\bx)$ cannot be
factorized as in Eq.~\ref{disp9} for $x>0$.

In order to write the equation of transformation of the PS let us
remind the basic relation between the CF and the PS of a spatially
stationary stochastic process:
\be
\int\!\!\int d^dx d^dy\; e^{-i(\bk\cdot\bx+\bq\cdot\by)}
C(\bx-\by)=(2\pi)^d \delta(\bk+\bq) S(\bk)\,.
\label{disp11}
\ee
We also recall that by definition:
\[
\left<\overline{n(\bx)n(\by)}\right>=n_0^2
+C(\bx-\by)\,.
\]
Furthermore, we define the functions
$\hat{f}(\bk_1,\bk_2;\bx)$ and
$F(\bk_1,\bk_2;\bq)$ respectively by the following FTs:
\be
\label{disp13}
\hat{f}(\bk_1,\bk_2;\bx) =\int\!\!\int d^du d^dv\;
e^{-i(\bk_1\cdot\bu+\bk_2\cdot\bv)}
f(\bu,\bv;\bx)\,,
\ee
\be
F(\bk_1,\bk_2;\bq)=\int d^dx \; 
e^{-i\bq\cdot\bx}\hat{f}(\bk_1,\bk_2;\bx)\,.
\label{disp13b}
\ee
The function $\hat{f}(\bk_1,\bk_2;\bx)$ is simply
the characteristic function of the joint two-displacement PDF.  By
definition $\hat{f}(\bk_1,\bk_2;\bx)$ satisfies the
following limit conditions
\[\hat{f}(0,0;\bx)=1\;\;\mbox{for any}\;\bx\,,\] 
and 
\[\hat{f}(0,\bk;\bx)=\hat{f}(\bk,0;\bx)=\hat p(\bk)\;\;
\mbox{for any}\; x>0\,.\]

By using Eq.~\ref{disp7bb} and Eqs.~\ref{disp11}-\ref{disp13b}, we can
write:
\bea
\label{disp14}
&&S(\bk)= n_0\left(1- \frac{1}{(2\pi)^d}
\int d^dq\,F(\bk,-\bk;\bq)\right)+\\
&&\int d^dx\, e^{-i\bk\cdot\bx}\hat{f}(\bk,-\bk;\bx)
\left(n_0^2+C_{in}(\bx)\right)-(2\pi)^dn_0^2\delta(\bk)\,.
\nonumber
\eea
Note that $\frac{1}{(2\pi)^d}\int d^dq\,F(\bk,-\bk;\bq)$
must be carefully handled. In fact, due to the properties of the
inversion of the Fourier transform, it cannot be substituted directly
by $\hat{f}(\bk,-\bk;0)$ if $f(\bu,\bv;\bx)$ is
discontinuous at $x=0$ and continuous anywhere else as in the case of
uncorrelated displacements presented above. Thus it must be
understood as
\be 
\frac{1}{(2\pi)^d}\int d^dq\,F(\bk,-\bk;\bq)=
\lim_{x\rightarrow 0} \hat{f}(\bk,-\bk;\bx)\,.
\label{FF}
\ee
More specifically, if $f(\bu,\bv;\bx)$ is continuous at $x=0$, i.e.,
\[\lim_{x\rightarrow 0} f(\bu,\bv;\bx)=
f(\bu,\bv;0)\equiv \delta(\bu-\bv)p(\bu)\,,\]
then we have
\[\frac{1}{(2\pi)^d}\int d^dq\,F(\bk,-\bk;\bq)=1\,.\]
This condition is valid in all the cases in which the stochastic
displacement field is a real continuous correlated stochastic process
(see below the Gaussian example). In fact, in this case it is possible
to prove a theorem \cite{gnedenko} stating that the two-displacement
correlation function is continuous everywhere, being equal to the
one-displacement variance $\overline{u^2}-\overline{u}^2$ at $x=0$.
Instead, in the case of an uncorrelated stochastic displacement field
this is no more true (it is not a continuous stochastic process),
and, $f(\bu,\bv;\bx)$ is discontinuous at $x=0$ as shown
by Eq.~\ref{disp9}. This together with Eq.~\ref{FF} gives for this
case
\[\frac{1}{(2\pi)^d}\int d^dq\,F(\bk,-\bk;\bq)=
\left|\tilde p(\bk)\right|^2\,.\]
With this prescriptions it is simple to recover Eq.~\ref{disp10} from 
Eq.~\ref{disp14} in the case of uncorrelated displacements.

Instead, in the present case of a stationary correlated continuous
stochastic displacement field, Eq.~\ref{disp14} can be rewritten as
\bea
S(\bk)&=&\int d^dx\; e^{-i\bk\cdot\bx}
\hat{f}(\bk,-\bk;\bx)  \left[n_0^2+C_{in}(\bx)\right]\nonumber\\
&-&(2\pi)^d n_0^2\,\delta(\bk)\,,
\label{disp15}
\eea
or equivalently
\bea
&&S(\bk)=n_0^2 F(\bk,-\bk;\bk)+\\
&&\frac{1}{(2\pi)^d}\int d^dq\, F(\bk,-\bk;\bq)
S_{in}(\bk-\bq)-(2\pi)^dn_0^2\delta(\bk)\,.\nonumber
\label{disp15b}
\eea
Equations \ref{disp15} can be further simplified by noticing that in
the case of spatial statistical stationarity of both the initial
point-particle distribution and of the displacement field, the PS of
the final particle distribution will not depend separately on the
couple of displacements $\bu$ and $\bv$ applied at two points
separated by the separation vector $\bx$, but on the relative
displacement ${\bf w}=\bu-\bv$.  In fact let us call
$\phi({\bf w};\bx)$ the PDF that two points, separated by the
separation vector $\bx$, perform a relative displacement
${\bf w}$. By definition
\be
\phi({\bf w};\bx)=\int\int d^du\,d^dv\,f(\bu,\bv;\bx)\,
\delta({\bf w}-\bu+\bv)\,.
\label{disp15c}
\ee
If we take the FT of Eq.~\ref{disp15c} with respect to ${\bf w}$, we have
\[
\hat \phi(\bk;\bx)=\hat f(\bk,-\bk;\bx)\,,
\]
where $\hat \phi(\bk;\bx)=FT[\phi({\bf w};\bx)]$.
Therefore Eq.~\ref{disp15} can be rewritten again
\be
S(\bk)=\int d^dx\, 
e^{-i\bk\cdot\bx}\hat \phi(\bk;\bx)
\left[n_0^2+C_{in}(\bx)\right]-(2\pi)^dn_0^2\delta(\bk)\,.
\label{disp15e}
\ee 
By using the PDF $\phi({\bf w};\bx)$, Eq.~\ref{disp7b} can be
rewritten in a very intuitive form. This is done by noticing that for
a generic stationary point process $n(\bx)$ the quantity
$\left<n(\bx)n(\by)\right>/n_0$ (where $\left<...\right>$ is as usual
the average over the considered ensemble of configurations $n(\bx)$)
represents the average {\em conditional} density
\cite{libro} of particles seen by a generic particle of the system at
a separation $\bx-\by$ from it. By calling
$\Gamma_{in}(\bx-\by)=\left<n_{in}(\bx)n_{in}(\by)\right>/n_0$ and
$\Gamma_f(\bx-\by)=\left<\overline{n(\bx)n(\by)}\right>/n_0$ the
conditional densities respectively before and after the application of
the displacement field, Eq.~\ref{disp7b} transforms into: 
\[
\Gamma_f(\bx)=\int d^dx'\;\Gamma_{in}(\bx')\,\phi(\bx-\bx';\bx')\,.
\]

Very often the CF is defined in a dimensionless way dividing
$C(\bx)$ by $n_0^2$, i.e., it is redefined as 
\[
\xi(\bx)=\frac{C(\bx)}{n_0^2}=\frac{\delta(\bx)}{n_0}+h(\bx)\,,
\]
which we call normalized covariance function (NCF).
Consequently, also the PS is redefined as the FT of $\xi(\bx)$:
\[
P(\bk)=FT[\xi(\bx)]=\frac{S(\bk)}{n_0^2}=\frac{1}{n_0}
+\hat h(\bk)\,.
\]
It is simple to verify that if all the particles of the distribution
have the same {\em non-unitary} mass $m>0$, i.e., if the microscopic
mass density is
$\rho(\bx)=m\,n(\bx)=m\sum_i\delta(\bx-\bx_i)$ with 
$\rho_0=m\,n_0$, the
NCF $\xi(\bx)$ of the microscopic {\em mass} density
$\rho(\bx)$ is equal to that of the microscopic {\em number}
density $n(\bx)$ and then is independent of $m$.
For these rescaled quantities Eq.~\ref{disp15e} can be rewritten as
\be
P(\bk)=\int d^dx\, 
e^{-i\bk\cdot\bx}\hat \phi(\bk;\bx)
\left[1+\xi_{in}(\bx)\right]-(2\pi)^d\delta(\bk)\,,
\label{disp15e-bis}
\ee 
It is important to note that, while $\xi_{in}(\bx)$ depends on $n_0$
at least through its diagonal part $\delta(\bx)/n_0$, the
two-displacement $f(\bu,\bv;\bx)$, and therefore $\hat \phi(\bk;\bx)$,
in our hypothesis, are in general supposed not to be (unless for
particular choices of its {\em momenta}). Therefore, differently from
the case of uncorrelated displacements, both Eqs.~\ref{disp15} and
\ref{disp15b} can be divided into two parts, one dependent on $n_0$,
and the other independent of it. This is a very important point
because spatial distributions of particles with equal masses $m$ are
often used in numerical simulations as a discrete representation of
continuous stochastic mass density fields.  Consequently, in
Eq.~\ref{disp15} and Eq.~\ref{disp15b} there is a part depending on
the discretization process and another part independent of it. This
aspect is particularly important in the context of the
gravitational $n-$body simulations in which, as above mentioned, the
matter density field is usually represented, for numerical reasons,
through a more or less dense distribution of particles with the same
mass $m$ \cite{pre-initial} so that $m n_0=\rho_0$ with $\rho_0$ put
equal to the average mass density of the continuous model.

\subsection{Small $k$ expansion and large scale behavior - II}
\label{small-k2}

Starting from Eq.~\ref{disp15e-bis}, one can write a simple formula
for the small $k$ (i.e., large scale) behavior of the final PS of the
particle distribution after the application of the displacement field.
First of all let us study the small $k$ behavior of $\hat
\phi(\bk;\bx)=\hat f(\bk,-\bk;\bx)$. Since it is the
FT of $\phi({\bf w};\bx)$, if $ G_{\mu\nu}({\bf 0})<+\infty$ (implying that
the average value of $w^2$ is finite at any $\bx$), we can write
\be
\hat \phi(\bk;\bx)=1-i\bk\cdot\overline{{\bf w}(\bx)}-
\frac{\overline{\left[\bk\cdot{\bf w}(\bx)\right]^2}}{2}+o(k^2)\,,
\label{s1}
\ee
where ${\bf w}(\bx)$ is the relative displacement
between two particles initially separated by a vector distance
$\bx$.
Let us suppose $\overline{{\bf w}(\bx)}=0$ which is automatic in the
case of statistical invariance for space inversion or any rotation 
(isotropy).
Moreover, using the definition of $G_{\mu\nu}({\bf x})$ we can write
\[\overline{\left[\bk\cdot{\bf w}(\bx)\right]^2}=
2\sum_{\mu,\nu}^{1,d}k^{(\mu)} k^{(\nu)}\left[G_{\mu\nu}({\bf 0})- G_{\mu
\nu}(\bx)\right]\,,\] where we have supposed also $G_{\mu
\nu}(\bx)=G_{\mu \nu}(-\bx)$.  Using this expression and
Eq.~\ref{s1} in Eq.~\ref{disp15e-bis}, we obtain at small enough $k$
\bea
\label{s2}
&&P(\bk)\simeq P_{in}(\bk)\left[1-
\sum_{\mu,\nu}^{1,d}k^{(\mu)} k^{(\nu)} G_{\mu\nu}({\bf 0})\right]
\\
&&+\sum_{\mu,\nu}^{1,d}k^{(\mu)} k^{(\nu)} \left[\hat G_{\mu\nu}(\bk)+
\int \frac{d^dq}{(2\pi)^d}\, \hat G_{\mu \nu}(\bq) 
P_{in}(\bk-\bq)\right]\,,
\nonumber
\eea
where $\hat G_{\mu \nu}(\bk)=FT\left[G_{\mu\nu}(\bx)\right]$ is the 
power spectrum matrix of the displacement field.
Note that, since $G_{\mu\nu}({\bf 0})=\int \frac{d^dq}{(2\pi)^d}
\hat G_{\mu \nu}(\bq)$, Eq.~\ref{s2} can be also rewritten as:
\bea
&&P(\bk)\simeq P_{in}(\bk)+\sum_{\mu,\nu}^{1,d}k^{(\mu)} k^{(\nu)} \left\{\hat G_{\mu\nu}(\bk)\right.
\nonumber\\
&&\left.+
\int \frac{d^dq}{(2\pi)^d}\, \hat G_{\mu \nu}(\bq) 
\left[P_{in}(\bk-\bq)-P_{in}(\bk)\right]\right\}\,,
\nonumber
\eea
which is useful in particular in the case in which the initial 
particle configuration is the stationary Poisson one for which 
$P_{in}(\bk)\equiv 1/n_0$ and the last term vanishes.

Depending on the small $k$ properties of $P_{in}(\bk)$ and $\hat
G_{\mu \nu}(\bk)$, the small $k$ behavior of $P(\bk)$ will
change.  Note that Eq.~\ref{s2} is valid only if
$G_{\mu\nu}({\bf 0})<+\infty$.  In the case in which
$G_{\mu\nu}({\bf 0})=+\infty$ then the singular part of the small $k$
expansion of $\hat\phi(\bk;\bx)$ has to be considered in a similar
way to the case of uncorrelated displacements given by
Eqs.~\ref{as1-2} and \ref{as1-3}.  A very particular and important
case of Eq.~\ref{s2} is given when perpendicular displacements are
not correlated at any $\bx$. This means that
$G_{\mu\nu}(\bx)=\delta_{\mu\nu}G(\bx)$ and $\hat G_{\mu
\nu}(\bk)=\delta_{\mu\nu}\hat G(\bk)$ (with $\delta_{\mu\nu}$
the Kronecher delta). If the displacement field is also isotropic,
$\hat G(\bk)$ depends only on $k$ (and consequently $G(\bx)$ on $x$),
so that the PS matrix with elements $\hat G_{\mu \nu}(\bk)$ is
invariant under any spatial rotation.

In this case Eq.~\ref{s2} can be rewritten as
\bea
\label{s3}
&&P(\bk)\simeq P_{in}(\bk)\left[1-G({\bf 0})k^2\right]\\
&&+k^2\left[\hat G(\bk)+\int \frac{d^dq}{(2\pi)^d}\, \hat G(\bq)\, 
P_{in}(\bk-\bq)\right]
\nonumber
\eea
Similarly to Eq.~\ref{s2}, also Eq.~\ref{s3} can be re-expressed as 
follows:
\bea
\label{s3b}
&&P(\bk)\simeq P_{in}(\bk)+k^2\left\{\hat G(\bk)\right.\\
&&\left.+\int \frac{d^dq}{(2\pi)^d}\, \hat G(\bq)\, 
[P_{in}(\bk-\bq)-P_{in}(\bk)]\right\}
\nonumber
\eea

Equations \ref{s3} and \ref{s3b} are very useful because to show very
clearly how effective a displacement field can be in ``injecting''
large scale correlations into a given particle distribution.  As
better clarified below, a central role is played by the spatial
dimension $d$.  Let us suppose that at small $k$ we have
$P_{in}(\bk)\sim k^{\alpha}$ and that $\hat G(\bk)\sim k^{\beta}$ (as
already shown $\alpha,\beta>-d$). Note that
\[ \int \frac{d^dq}{(2\pi)^d}\, \hat G(\bq)\, 
P_{in}(\bk-\bq)=\int d^dx\,e^{-i\bk\cdot \bx}
G(\bx)\xi_{in}(\bx)\,,\] and $\xi_{in}(\bx)$ goes to zero
at large $x$. Hence, when $\beta< 0$, in Eq.~\ref{s3}, the term $k^2\hat
G(\bk)$ is more effective than $k^2\int \frac{d^dq}{(2\pi)^d}\,
\hat G(\bq)P_{in}(\bk-\bq)$ 
(i.e., the latter has an exponent larger than
$\beta+2$).  For $\beta=0$, apart from the very particular
case in which $\int \frac{d^dq}{(2\pi)^d}\,
\hat G(\bq)P_{in}(\bq)=0$, both terms are of the same order at
small $k$, i.e., proportional to $k^2$.  Instead for $\beta>0$, apart
again the previous particular choices of $\hat G(\bk)$ in relation to
$P_{in}(k)$, $k^2\int \frac{d^dq}{(2\pi)^d}\,\hat
G(\bq)P_{in}(\bk-\bq)\sim k^2$ prevails on $k^2\hat G(\bk)\sim
k^{\beta+2}$ at sufficiently small $k$.

We can therefore conclude that
\begin{itemize}
\item $\beta<0$: if $\alpha<\beta+2$ the displacement field is
ineffective in changing large scale two-point correlations between
particles. In fact the small $k$ leading term of the final PS
$P(\bk)$ is the initial one $P_{in}(\bk)$.  For what concerns
perturbations to this leading term, it is simple to show that, if
$\alpha<\beta$ the main perturbation to the leading term due to the
displacements is $-G({\bf 0})k^2P_{in}(\bk)\sim k^{\alpha+2}$,
while if $\beta<\alpha<\beta+2$ it is $k^2\hat G(\bk)\sim
k^{\beta+2}$ (if $\alpha=\beta$ the two terms are of the same order).

If instead $\alpha>\beta+2$, the displacement field completely
determines the new large scale correlation properties of the particle
distribution being $k^2\hat G(\bk)\sim k^{\beta+2}$ now the leading
term of $P(\bk)$.  Since $\beta>-d$, the limit ``most critical''
behavior of $P(\bk)$ which can be reached is $k^{2-d}$.  Note that for
$d\ge 3$ long range non-integrable and mainly positive (i.e.,
critical) two-point correlations can be introduced in the system by
the action only of displacements having finite variance $G({\bf
0})$. Accordingly the (unreachable) limit decaying behavior for the NCF at
large $x$ is given by $\xi(\bx)\sim x^{-2}$ which is not integrable
and long range.  For example in $d=3$ one can start from a completely
uncorrelated Poisson particle distribution, and, applying finite
displacements (i.e., with a finite average squared value) but with
long range correlations, to obtain a particle distribution with a
covariance function similar to those found at the critical point of a
second order phase transition.  For $d\le 2$ the exponent $2-d\ge 0$,
hence $\xi(\bx)$ decays faster than $x^{-d}$ and correlations are not
``critical'', but integrable (i.e., for our purpose, short range).

Finally, for $-2<\beta<0$ and $\alpha>\beta+2$ and any $d$, even
though the leading term of $P(\bk)$ is due to the displacement field
which is characterized by long range correlations, the final particle
distribution remains superhomogeneous (see Sec.~\ref{basic}).

\item $\beta=0$. If $\alpha<2$ the leading term of the final PS $P(\bk)$
is again $P_{in}(\bk)$. The displacement field introduces only
higher order perturbations: (1) $-G({\bf 0})P_{in}(\bk)k^2\sim
k^{\alpha+2}$, and (2) $k^2\left[\hat G(\bk)+\int
\frac{d^dq}{(2\pi)^d}\, \hat G(\bq)\,
P_{in}(\bk-\bq)\right]\sim k^2$.  The former is the most important
perturbative term for $\alpha<0$ (``critical'' initial condition),
while the latter is when $0<\alpha<2$. For $\alpha=0$ (1) and (2) are
in general of the same order and both contribute to the main perturbation
to $P_{in}(\bk)$. However, if the initial particle
configuration is a stationary Poisson one, $P_{in}(\bk)\equiv 1/n_0$
for any $\bk$. Consequently, as shown explicitly by Eq.~\ref{s3b}, the
main perturbation to $P_{in}(\bk)$ is only $k^2\hat G(\bk)$, as the 
other terms in Eq.~\ref{s3b} cancel one each other. We show this better 
in the following through the example of the Gaussian displacement field.

Instead, if $\alpha>2$ again the small $k$ behavior of the final PS is
completely different from the initial one and is determined by the
displacement field.  In fact now the leading term becomes
$k^2\left[\hat G(\bk)+\int
\frac{d^dq}{(2\pi)^d}\, \hat G(\bq)\,
P_{in}(\bk-\bq)\right]\sim k^2$. However the system after the action
of the displacements is still in the superhomogeneous class of
point processes in any $d$.

\item $\beta>0$. 
If $\int \frac{d^dq}{(2\pi)^d}\,\hat G(\bq)P_{in}(\bq)>0$, this case
is similar to the case $\beta=0$ with the difference that when
$\alpha>2$ the main term introduced by the displacement field is only
$k^2\int
\frac{d^dq}{(2\pi)^d}\, \hat G(\bq)\, P_{in}(\bk-\bq)\sim k^2$.
However if $\int\frac{d^dq}{(2\pi)^d}\, \hat G(\bq)\, P_{in}(\bq)=0$,
which corresponds to a peculiar choice of the dependence of $\hat
G(\bq)$ on $P_{in}(\bq)$, the term $k^2\hat G(\bk)$ can again be
important. This case will be analyzed in more detail in the next
section when the case of a shuffled lattice with correlated Gaussian
displacements is analyzed.  For the moment we notice only that this
case is quite difficult to be realized because of the condition given
by the Wiener-Khinchin theorem \cite{torquato-book} which states that
both $P_{in}(\bk)$ and $\hat G(\bk)$ must be nonnegative at any $\bk$.
This means that it can be realized only if $P_{in}(\bk)$ is zero where
$\hat G(\bk)$ is not and vice-versa, i.e., non-overlapping supports.

Finally, note that when the initial particle configuration is a
regular lattice, the term $P_{in}(\bk)\left[1-G({\bf 0})k^2\right]$ in
Eq.~\ref{s3} is identically zero around $k=0$. Consequently, the small
$k$ PS of the final configuration is determined by the next
perturbation terms (see the next section about Gaussian
displacements).

\end{itemize}

It is important to notice that Eqs.~\ref{s2} and \ref{s3}
are quite more complex than the result obtained
in a {\em naive} way in Sec.~\ref{naive} of the paper by simply using
the continuity equation for the local conservation of mass which led 
basically to
\[P(k)=k^2\hat G(k)\,.\]
We see that respect to this simple approximation, even in the small
$k$ limit (i.e., large spatial scales) and finite displacements (i.e.,
finite $G({\bf 0})$), there can be important corrections coming both
from the granularity of the system, from the internal correlations of
the initial mass distribution, and from the interplay between these
correlations and those of the displacement field.

\section{Correlated Gaussian displacement field.} 
\label{gauss}

In this section the effect of a correlated displacement field on the
correlation properties of a given particle distribution is further
clarified through the discussion of a very important example: the
Gaussian displacement field. Its importance is
twofold: (1) its statistical properties are completely determined by
the knowledge of one and two-point correlation functions (i.e., mean
value $\overline{\bu}$ and correlation matrix $G_{\mu\nu}(\bx)$); (2)
in many applications (e.g., initial conditions of cosmological
$n-$body simulations) the Gaussian of perturbations (i.e., particle
displacements) arises as a natural hypothesis basically due to
arguments based on the {\em central limit theorem} \cite{gnedenko}.
Moreover, in order to clarify better all the concepts introduced in
the previous section, two explicit examples of application of a
Gaussian displacement field will be given: stationary Poissonian
and regular lattice initial conditions.

We treat in detail the particular case of a {\em one-dimensional} spatially
stationary point process $n_{in}(x)$ perturbed by a statistically
stationary Gaussian displacement field $u(x)$.  The
{\em probability density functional} ${\cal Q}[u(x)]$ giving the
statistical weight of any realization $u(x)$ of the field will have
the form \cite{libro}:
\be
{\cal Q}[u(x)]\sim \exp \left[-\frac{1}{2}\int\!\!
\int_{-\infty}^{+\infty} dx\,dy\,u(x)K(x-y)u(y)\right]\,,
\label{disp18}
\ee
where $K(x)=K(-x)$ is the positive definite correlation kernel of the
Gaussian displacement field. Without loss of generality we have put
$\overline{u}=0$. Clearly Eq.~\ref{disp18} is only formal as,
rigorously speaking, the normalization constant is infinite in the
continuum space and also in the infinite volume limit. However we write it
to make evident the meaning of Gaussianity for a
stochastic field.  For a more general account of Gaussian stochastic
fields see \cite{gardiner,adler}.

It is simple to show that the PS of the displacement field is given by
\[
\hat G(k)=\frac{1}{FT[K(x)]}\,.
\]
As we want $u(x)$ to be a well defined continuous stochastic process,
$K(x)$ is to be taken so that the Wiener-Khinchin theorem is satisfied
\cite{gnedenko}, i.e., $K(x)$ is such that $\hat G(k)\ge 0$ at all $k$
and integrable.  The displacement-displacement correlation function
will be thus given by
\[
G(x)\equiv\overline{u(x_0+x)u(x_0)}=FT^{-1}\left[\frac{1}{FT[K(x)]}\right]\,,
\]
which is a continuous function if $u(x)$ is a continuous
stochastic process \cite{gnedenko}. It is possible to show that the
joint two-displacement PDF can be written as:
\bea
\label{disp19}
f(u,v;x)=&&\frac{1}{2\pi\sqrt{G^2(0)-G^2(x)}}\times\\
&&\exp \left[-
\frac{G(0)(u^2+v^2)-2G(x)uv}{2(G^2(0)-G^2(x))} \right]\,,\nonumber
\eea
where $G(0)=\overline{u^2}\equiv \overline{v^2}<+\infty$.
It is also simple to verify that, in order to have this joint PDF 
well defined at all $x$, the correlation function $G(x)$ must satisfy 
the following constraint:
\[|G(x)|< G(0) \;\;\forall x\ne 0\,.\]
Note that if, as supposed, the Gaussian displacement field is a real
continuous correlated stationary stochastic process 
and, consequently $G(x)$ is a continuous
function, then $\lim_{x\rightarrow 0}G(x)=G(0)$.  This implies then
$\hat f(k,-k;0)=1$. If instead the displacement field is uncorrelated
(i.e., $G(x)=0$ for $x\ne 0$ but $G(0)>0$), then, as shown in
Sec.~\ref{indip}, $\hat f(k,-k;0)=\exp [-k^2G(0)]\equiv \left|\hat
p(k)\right|^2$ with $\hat p(k)=\exp [-k^2G(0)/2]$ being the
characteristic function of the one-displacement PDF.

In the case of truly continuous process, by taking the FT of
Eq.~\ref{disp19} both in $u$ and $v$, we find:
\be
\hat f(k,-k;x)\equiv \hat \phi(k; x)=e^{-k^2(G(0)-G(x))}\,.
\label{disp20}
\ee 
Therefore, by using Eq.~\ref{disp20}, the relation \ref{disp15}
between the PS of the particle distribution after the application of
the displacement field, and its initial correlation function
$\xi_{in}(x)$ will be 
\bea 
P(k)=&&e^{-k^2G(0)}\int_{-\infty}^{+\infty}
dx\,e^{-ikx+k^2G(x)} \left(1+\xi_{in}(x)\right)-\nonumber\\
&&2\pi\delta(k)\,.
\label{disp21}
\eea
Since at large $x$ both $G(x)$ and $\xi_{in}(x)$ converge to zero, the
FT in $x$ in Eq.~\ref{disp21} is not well defined and contains a Dirac
delta function contribution exactly compensating the last Dirac delta
function term in the same equation.  This can be made more clear in
the following way.  The Dirac delta function of Eq.~\ref{disp21} can
be transformed as:
\[2\pi\delta(k)=2\pi e^{-k^2G(0)}\delta(k)=
e^{-k^2G(0)}\int_{-\infty}^{+\infty} dx\,e^{-ikx}\,.\]
Using this relation, Eq.~\ref{disp21} can be rewritten in the
following form containing only well defined FTs:
\bea
P(k)=&&e^{-k^2G(0)}\left[\int_{-\infty}^{+\infty} dx\,e^{-ikx}
\left(e^{k^2G(x)}-1\right)+\right.\nonumber\\
&&\left.\int_{-\infty}^{+\infty} dx\,e^{-ikx+k^2G(x)}
\xi_{in}(x)\right]\,.
\label{disp22}
\eea

The small $k$ expansion of this formula can be obtained by simply
applying to this case Eq.~\ref{s3}.

As already shown above, in $d>1$ dimensions the more general form of
$f(\bu,\bv;\bx)$, even in the limited case of spatially
stationary Gaussian displacement field is more complex. In fact in
general not only parallel components of $\bu$ and $\bv$ can be
correlated, but also perpendicular components can be. This leads to
have a symmetric displacement-displacement correlation matrix
$G_{\mu\nu}(\bx)$.  Once the correlation matrix is given, it is
simple to show that, for a $d-$dimensional spatially stationary
Gaussian displacement field with zero average value, we can write:
\bea
&&\hat \phi(\bk;\bx)\equiv \hat f(\bk,-\bk;\bx)=\nonumber\\
&&\exp\left[-\sum_{\mu,\nu}^{1,d} k^{(\mu)}k^{(\nu)}
[G_{\mu\nu}({\bf 0})-G_{\mu\nu}(\bx)]\right]\,.\nonumber
\eea
However, in the case $G_{\mu\nu}(\bx)=\delta_{\mu\nu} G(x)$, 
the general features of
the effect of a Gaussian displacement field can be well summarized by
the above one-dimensional example.

\subsection{Gaussian displacement field applied to a 
Poisson particle distribution}
\label{poisson}

In this section we analyze in detail the effect of a Gaussian
displacement field as presented above on a one-dimensional Poisson particle
distribution. The extension to higher dimensions is straightforward.

Let us suppose to have a spatially stationary Poisson particle distribution 
with constant average number density $n_0>0$.
As aforementioned in this case the NCF is
\[\xi_{in}(x)=\frac{\delta(x)}{n_0}\,,\]
i.e., $h_{in}(x)=0$ and $P_{in}(k)=\frac{1}{n_0}$.
By substituting this initial NCF into Eq.~\ref{disp22}, one finds:
\[
P(k)=\frac{1}{n_0}+e^{-k^2G(0)}\int_{-\infty}^{\infty}
dx\,e^{-ikx}\left(e^{-k^2G(x)}-1\right)\,,
\]
which, in the long wave-lengths limit, behaves as
\be
P(k)\simeq\frac{1}{n_0}+k^2\left[\hat G(k) -k^2G(0)\right]\,.
\label{poi-corr-2}
\ee
Since in $d=1$, $\hat G(k)\sim k^{\beta}$ with $\beta>-1$, we see that
at small $k$ the leading term is $1/n_0\equiv P_{in}(k)$ and the
perturbations introduced by the displacements are at most of order
$k$. However, as aforementioned, in $d>2$ ``critical'' and
dominating perturbations at small $k$ can be introduced by the
displacements. 
On the other hand, as clear from Eq.~\ref{poi-corr-2}, we have a
perturbation of order $k^4$ at small $k$ due to the finite variance of 
the displacements $G(0)$ which prevails on the perturbation coming
from displacement-displacement correlations (i.e., from the term containing
$\hat G(k)$ in Eq.~\ref{poi-corr-2}) if $\beta>2$. 
We ultimately observe that for any $d$ it is not possible
to transform a Poisson particle distribution into a
``superhomogeneous'' one by the only action of the displacements. In
fact, even though correlated, the displacements are stochastic
perturbations and, consequently, cannot increase the level of large
scale order of the system.

\subsection{Shuffled lattice with correlated Gaussian displacements}
\label{shuffled2}

In this subsection we analyze the effect of the same
Gaussian displacement field as above on a one dimensional lattice
(i.e., a regular chain of unit mass particles) with lattice spacing $l$
(i.e., $n_0=1/l$). Again we focus our attention on the large
scales ($k\rightarrow 0$ limit).  By taking the inverse
FT of Eq.~\ref{sh-lat1}, the initial NCF in this case is
\[
\xi_{in}(x)=l\sum_{n=-\infty}^{+\infty}\delta(x-nl)-1\,.
\]
Therefore Eq.~\ref{disp21} can be rewritten as
\be
P(k)=le^{-k^2G(0)}\sum_{n=-\infty}^{+\infty}e^{-iknl+k^2G(nl)} 
-2\pi\delta(k)\,.
\label{csh-ps}
\ee
The series in Eq.~\ref{csh-ps} is not well defined because
its argument does not converge to zero for $|n|\rightarrow \infty$.
By using for $2\pi\delta(k)$ the chain of identities
\bea
&&2\pi\delta(k) \equiv 2\pi e^{-k^2G(0)}\times\left[\sum_{m=-\infty}^{+\infty}
\delta\left(k-{2\pi \frac{m}{l}}\right)\right.
\nonumber\\
&&\left.-
\sum_{m\ne 0}^{-\infty,+\infty}\delta\left(k-{2\pi \frac{m}{l} }\right)
\right]\equiv l e^{-k^2G(0)}\times\sum_{n=-\infty}^{+\infty}e^{-iknl}
\nonumber\\
&&- 2\pi e^{-k^2G(0)}\times\sum_{m\ne 0}^{-\infty,+\infty}
\delta\left(k-{2\pi \frac{m}{l} }\right)\,,\nonumber
\eea
we can rewrite Eq.~\ref{csh-ps} as
\bea
&&P(k)=le^{-k^2G(0)}\sum_{n=-\infty}^{+\infty}e^{-iknl}\left[e^{k^2G(nl)}
-1\right]\nonumber \\
&&+\,2\pi e^{-k^2G(0)}\sum_{m\ne 0}^{-\infty,+\infty}
\delta\left(k-{2\pi \frac{m}{l} }\right)\,.
\label{sh-lat2-b}
\eea The first sum is a well defined series and gives in general the
smooth part of $P(k)$ converging to $1/n_0$ at large $k$. Instead the
second one gives the singular delta-like Bragg peaks contribution to
$P(k)$ due to the initial perfect lattice distribution, each peak
being modulated by an amplitude $e^{-k^2G(0)}$, rapidly decreasing to
zero with increasing the position $k$ of the peak.  In $d>1$ it
is simple to obtain a very similar formula.

\bef[tbp]
\includegraphics[height=6cm,width=8cm,angle=0]{fig5.eps}
\vspace{0.2cm}
\caption{The figure presents the contribution to Eq.~\ref{sh-lat2-b} coming
from the first sum (i.e., excluding the delta-like Bragg peaks) to the
PS $P(k)$ of a $1d$ ``shuffled lattice'' obtained by applying a
correlated Gaussian displacement field with $\hat G(k)=\exp(-|k|/k_0)$
with $k_0=1/4$ (hence $G(0)=\overline{u^2}\simeq 0.08$) to a regular
chain of particles with unitary lattice spacing ($l=1$).  Both the
theoretical prediction given by Eq.~\ref{sh-lat2-b} and the numerical result
obtained by a direct simulation with $10^4$ particles are given.  The
agreement is excellent. Note that since in this case $\hat G_D(0)>0$
and finite we have that $P(k)\sim k^2$ at sufficiently small $k$.
\label{corr-k2}}
\eef
Since this second contribution, due to the initial NCF, is exactly
zero at all orders in the region $-2\pi/l < k< 2\pi/l$, the small $k$
behavior of $P(k)$ is completely determined by the first sum. This
situation is very different from the case of an initial Poisson
particle configuration. In fact, as shown above, in this second case
the small $k$ properties of the PS are determined mainly by the
displacement field only if it produces ``critical'' large scale
correlations (which is possible only in $d>2$).  Instead, in the
present case, due to the particular properties of the lattice PS, it
is always the displacement field to determine the large scale
correlation properties, i.e., the small $k$ PS.
We now analyze in detail the small $k$ behavior of $P(k)$ in 
Eq.~\ref{sh-lat2-b} with a particular attention to what kind of
superhomogeneous distributions we can obtain after the application of
the displacements.  
By keeping only the most important terms of 
Eq.~\ref{sh-lat2-b} at small $k$, we can write: 
\be P(k)
\simeq
e^{-k^2G(0)}k^2\hat G_D(k)+\frac{k^4}{2}[\hat G^{(2)}_D(k)-2G(0)\hat G_D(k)]\,,
\label{sh-lat3-b}
\ee 
plus higher order terms.  In Eq.~\ref{sh-lat3-b} $\hat G_D(k)$ is
the discretized Fourier integral of $G(x)$ 
with finite integration element given by
the lattice cell size $l$, i.e.,  
\be \hat
G_D(k)=\sum_{n=-\infty}^{+\infty}l e^{-iknl}G(nl)\,,
\label{sh-lat4-b}
\ee 
and analogously
\[
\hat G^{(2)}_D(k)=
\sum_{n=-\infty}^{+\infty}l e^{-iknl}[G(nl)]^2\,.
\]

Note that, due to the discretization of the Fourier integral in a sum
in Eq.~\ref{sh-lat4-b}, $\hat G_D(k)$ is not exactly equal to $\hat
G(k)$. This difference depends on how smooth and constant is $G(x)$ on
the length scale $l$ and vanishes when $l$ goes to zero.  However in
general if $\hat G(k)\sim k^\alpha$ at small $k$ with $\alpha\le 0$
then also $\hat G_D(k)\sim k^\alpha$.  Instead if $\alpha>0$, and then 
$\hat G(0)=0$, we can have $\hat G_D(0)>0$ (but
vanishing as $k^\alpha$ when $l\rightarrow 0$) as the effect of the
discretization of the Fourier transform.  Another important
observation about Eq.~\ref{sh-lat3-b} is that, as $[G(x)]^2\ge 0$ at
all $x$, then $\hat G^{(2)}_D(0)>0$.  Moreover, as $[G(x)]^2$ decays
faster than $G(x)$ at large $x$ then, if $\hat G_D(0)$, is finite also
$\hat G^{(2)}_D(0)$ is.

These observations are very important in order to determine which kind
of point processes can be realized by perturbing a lattice through a
correlated stochastic displacement field.  By the previous
considerations, it is straightforward to see that the realization from
a lattice of a stochastic particle distribution with $P(k)\sim
k^\beta$ at small $k$ with $\beta\le 2$ ($1< \beta\le 2$ in $d=1$, and
$2-d< \beta\le 2$ in general $d$ dimensions) is a very simple task: it
is enough to take a Gaussian displacement field such that $\hat
G(k)\sim k^\alpha$ with $\alpha=\beta-2\le 0$ (see Figs.~\ref{corr-k2}
and \ref{2-corr-k}).  Instead, obtaining a $P(k)$ with $\beta>2$ is a
more difficult task. First of all one has to take a $G(x)$ such that
$\hat G_D(0)=0$. It is possible to see that this requirement is a sort
of stochastic expression of the total momentum conservation law in the
system, that is, a conservation ``in average'' \cite{k4}.  Moreover in
order to have $2<\beta<4$ one has to have $\hat G_D(k)\sim
k^{\beta-2}$ at small $k$.  If instead $\hat G_D(k)\sim k^{\alpha}$
with $\alpha \ge 2$ then $\beta=4$ is obtained in all cases.  The case
$\beta>4$ is not permitted. This means that, by perturbing a lattice
with a (Gaussian) stochastic displacement field, the most
``superhomogeneous'' realizable point process has a PS $P(k)\sim k^4$
at small $k$, e.g., $k^6$ is forbidden. This is a strong
limitation. In fact, as aforementioned, from one side the PS of the
lattice can be considered as behaving $\sim k^\infty$ around $k=0$ (it
is completely flat).  From the other side we have just shown that an
exponent $\beta>4$ of $P(k)$ is never realizable when one applies to
any initial particle distribution a stochastic displacement field.
This means that independently of the kind of the displacement field,
as long as a smooth $\hat G(k)$ around $k=0$ is considered, there is a
``minimal level of disorder'' injected in the system measured by the
exponent $4$ in the small $k$ behavior of $P(k)$. This shows how
difficult is for example to build a spatially stationary stochastic
point process such that $P(k)\sim k^6$ at small $k$.
\bef[tbp]
\includegraphics[height=6cm,width=8cm,angle=0]{fig6.eps}
\vspace{0.2cm}
\caption{Power spectrum $P(k)$ 
(including also the Bragg peaks contribution) of a shuffled lattice
similar to the previous figure but now with $\hat
G(k)=A\exp(-k/k_c)/(k+a)$ with $k_c=5$, $a=10^{-4}$, and $A=1/20$
(hence $G(0)=\overline{u^2}=0.13$). The agreement between the
theoretical prediction and the numerical result is again excellent.
Note that now, as $\hat G(k)\sim k^{-1}$ at small $k$ (but larger than
$a$), the PS $P(k)\sim k$ in the same range.
\label{2-corr-k}}
\eef

It is possible to extend all these results to higher dimension and to
non-Gaussian displacement field, but we think that the example of the
one-dimensional Gaussian displacement field is enough to enlighten the
effects and the limitations of a general stochastic and correlated
displacement field on a given point process statistically independent 
of it.

\bef
\includegraphics[height=7.5cm,width=9cm,angle=0]{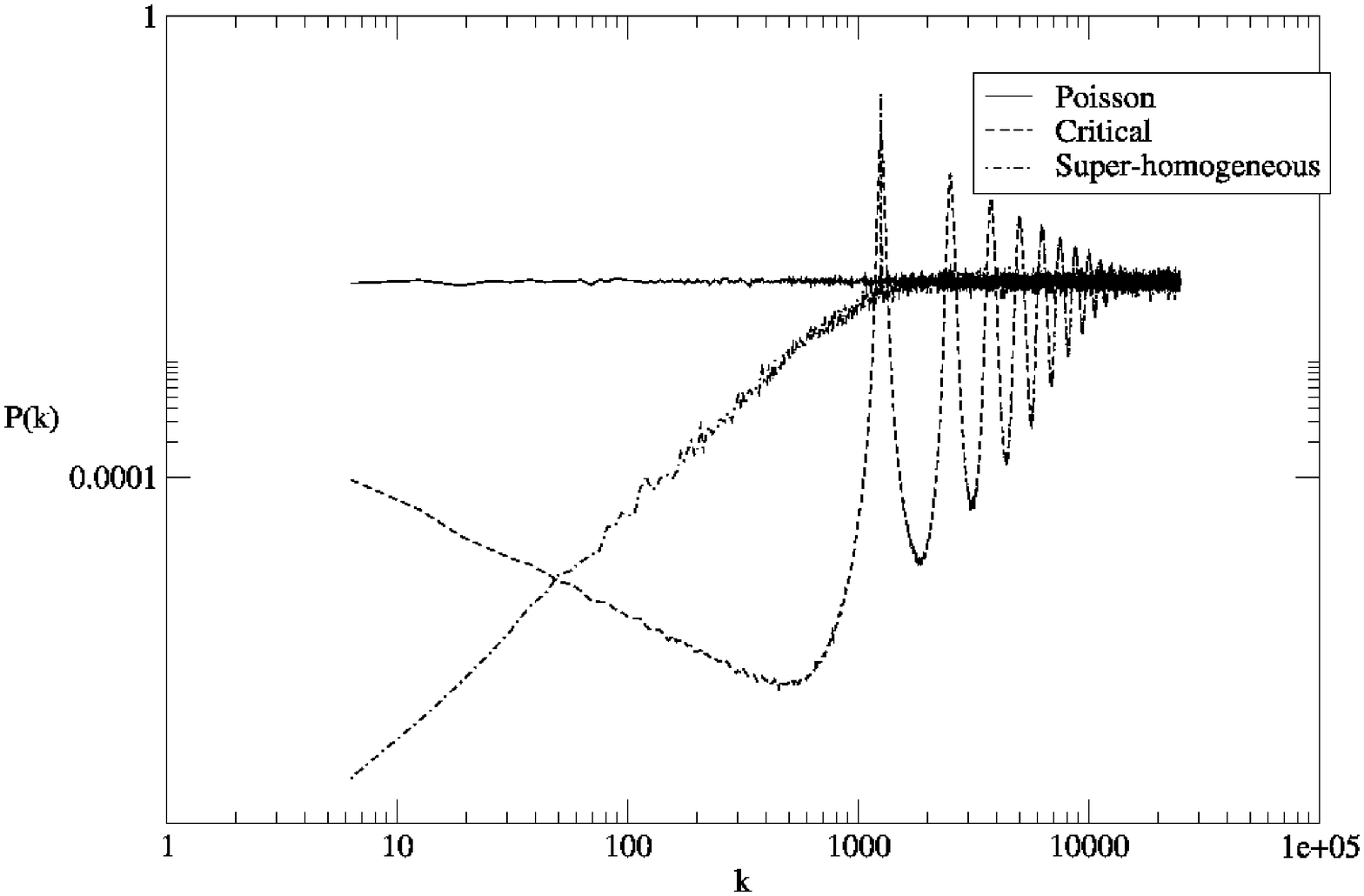}
\caption{The figure presents a comparison between the PSs of (i) a
Poisson point process, (ii) a shuffled lattice with uncorrelated
displacements with a finite variance, and (iii) a shuffled lattice
obtained by applying a correlated displacement field and showing a
``critical'' behavior at sufficiently small $k$.  In particular all
the particle distributions are in $d=1$ and have the same mean density
$n_0$ (i.e., the same specific volume $1/n_0$).  In the case (ii) the
finite variance $\overline{u^2}<+\infty$ implies $P(k)\sim k^2$ at
small $k$ (see Eq.~\ref{disp10}).  Instead in (iii) the applied
correlated displacement field for $a\ll k\ll 2\pi/n_0$ has $\hat
G(k)\sim k^{-4}$, and then, from Eq.~\ref{sh-lat3-b}, $P(k)\sim k^2$.
For $k<a$ (which is below the minimal $k$ of the figure) $\hat G(k)$
is cutoff to a constant value, in fact in any $d$ the PS of any proper
stochastic process has to satisfy $\lim_{k\rightarrow 0} k^dP(k)=0$.
\label{comparison}}
\eef

\section{Conclusions}

In this paper we have discussed the effect of a stochastic
displacement field on a spatial distribution of point-particles with
identical mass (i.e., point process) in the hypothesis of statistical
independence between the two.  In particular we have studied
rigorously the changes induced in the two-point correlation properties
of the particle distribution by the displacements.  In this way we
have seen that usual {\em naive} approaches to this problem leads to
approximations (which may not be appropriate in some important cases)
neglecting the contributions coming from either the initial correlation
properties of the particle distribution or the failure of the small
displacements approximation.

We have distinguished the two cases of displacement fields with 
and without spatial correlations, giving for both cases
the rigorous equations of transformation of the PS of the particle
distribution.  

Our main interest concerns a detailed study of the kind of large scale
correlations that the displacement field can ``inject'' into the
system. In particular we are interested to know if long range positive
spatial correlations in the system can be obtained by the application
of a suitable displacement field to an initially short range
correlated particle distribution. In this respect we have found that
in $d\ge 3$ it is possible to obtain particle distributions with long
range correlations (i.e., with a covariance function with a positively
diverging integral) by applying for instance either to a completely
uncorrelated homogeneous Poisson point process or to a regular lattice
of particles a Gaussian displacement field with sufficiently long
range displacement-displacement correlations independently on the
variance of the single displacement. This is a very important point,
in fact we can think to perturb a lattice with such a displacement
field with a mean squared displacement much smaller than the lattice
spacing, and, independently of this, to obtain strong large scale
typical fluctuations $\Delta N(R)$ of the number of points $N(R)$
contained in a randomly placed sphere of radius $R$, growing as
$\Delta N(R)\sim R^\alpha$ with $\alpha>d/2$, whereas $\alpha=d/2$ for
the Poisson distribution and $\alpha=(d-1)/2$ for the regular lattice.
Such particle distributions would look very similar respectively to
the original Poisson or lattice particle distribution at the small
scales (i.e., locally), but would show much larger (and more rapidly
growing with the distance) fluctuations beyond a sufficiently large scale.

Another related problem is to study how the long range order of a
regular lattice array of particles is perturbed by the action of the
applied displacement field. In particular we have studied between
which limits the perturbed lattice stays in the so-called {\em
superhomogeneous} class of point processes. That is, we have found the
limits in which the sub-Poissonian character of the mass (i.e. the
particles number) fluctuations on sufficiently large scales is
conserved.  Anyway we have also seen that any truly stochastic
displacement field always injects into the particle system at least a
minimal degree of disorder measured by the $\alpha=4$ of the exponent of
the final PS of the system $P(k)\sim k^\alpha$ at small $k$.

We think that all these results can be of very practical importance in
many physical and more largely scientific applications.  For instance,
this is the case of the so called $n$-body cosmological
simulations. In this case the superposition of a suitable stochastic
displacement field to a {\em pre-initial} particle distribution
(e.g., a lattice of identical particles) is the usual way to prepare
the {\em initial} conditions for numerical simulations to study the
problem of the dynamic gravitational clustering of the matter leading
to the formation of large scale structures (e.g., galaxies). These so
built initial conditions should represent the spectrum of small
primordial mass fluctuations predicted by theoretical models (e.g.,
CDM models).

\section{Acknowledgments}
The author thanks M. Joyce, F. Sylos Labini and L. Pietronero for
discussions and continuous collaborations on related
subjects. B. Marcos is warmly acknowledged for the fundamental help
given in doing Figs.~\ref{sl-k2}-\ref{2-corr-k}. 
The author thanks T. Baertschiger, S. Torquato and
P. Viot for interesting discussions.  Finally, the ``E. Fermi'' center
of Rome (Italy) is also acknowledged.


\begin{thebibliography}{99}
\bibitem {ziman} J. M. Ziman, {\em Principles of theory of solids}
$2$nd ed., (Cambridge Univ. press, Cambridge, 1972).
\bibitem {ashcroft} N. W. Ashcroft and N. D. Mermin, {\em Solid state 
physics}, (Saunders College, 1976).
\bibitem {groma} I. Groma, Phys. Rev. B, {\bf 56}, 5807 (1997).
\bibitem {hansen} J.-P.. Hansen and I. R. McDonald, {\em Theory of simple
liquids}, (Academic, New York, 1986).
\bibitem {radin} C. Radin, Notices Amer.Math.Soc., {\bf 42}, 26,
(1995); Chapter in {\em ``Geometry at Work''}, MAA Notes,
{\bf 53} (Math. Assoc. Amar., Washington, DC, 2000).
\bibitem {pee93} P. J. E. Peebles, {\it Principles Of Physical
Cosmology}, (Princeton University Press, 1993).
\bibitem {n-body1} S. D. M. White, {\em Lectures given at Les Houches} 
{\tt astro-ph/9410043} (1993).
\bibitem {cpu} C. L. Chan and A. K. Katsaggelos, IEEE Transactions of
Image Processing, {\bf 4}, 743 (1995); C. Su and L. Anand, {\tt
https://dspace.mit.edu/retrieve/3509/IMST015.pdf}.
\bibitem {renshaw} E. Renshaw, Biometrical Jour., {\bf 44}, 718 (2002).
\bibitem {daley} D. J. Daley and D. Vere-Jones, {\it An introduction
to the theory of point processes}, (Springer Verlag, Berlin, 1988).
\bibitem {kertscher} M. Kerscher, Phys. Rev. E, {\bf 64}, 056109 (2001).
\bibitem {torquato98} T. M. Truskett, S. Torquato, and P. G. Debenedetti,
Phys. Rev. E, {\bf 58}, 7369 (1998).
\bibitem {harrison-zeldovich} E. R. Harrison, Phys.Rev.D, {\bf 1},
2726 (1970); Ya. B. Zeldovich, Mon. Not. R. Acad. Soc., {\bf 160}, 1,
(1972).
\bibitem {pre-initial} G. Efstathiou, M. Davis, C. Frenk, and S. D. M. White,
Astrophys. J. Supp. Series, {\bf 57}, 241 (1985).
\bibitem {thierry1} T. Baertschiger  and F. Sylos Labini,
Europhys. Lett., {\bf 57}, 322 (2002).
\bibitem {BS95} P. Schneider and M. Bartelmann, Mon. Not. R. Astron. Soc.,
{\bf 273}, 475 (1995).
\bibitem {zeldovich-app} Ya. B. Zeldovich, Astron. \& Astrophys, {\bf 5},
84 (1970).
\bibitem {shuf-lat} A. Gabrielli, M. Joyce, and F. Sylos Labini,
Phys. Rev. D, {\bf 65}, 083523 (2002).
\bibitem {lebo} A. Gabrielli, B. Jancovici, M. Joyce, J. L. Lebowitz,
L. Pietronero, and F. Sylos Labini, Phys.Rev. D, {\bf 67},
043406,(2003).
\bibitem {thierry2} T. Baertschiger,  M. Joyce \& F. Sylos Labini,
Astrophys. J. Lett., {\bf 581}, L63 (2002).
\bibitem {gnedenko} B. Gnedenko, {\it The theory of probability},
(Mir Publishers, Moscow, 1975).
\bibitem {torquato-book} S. Torquato, {\it Random Heterogeneous
Materials}, Springer-Verlag, Berlin, 2002).
\bibitem {io-torquato} A. Gabrielli and S. Torquato, in press on Phys. 
Rev. E; {\tt cond-mat/0405115}.
\bibitem {torquato-real} S. Torquato and F. H. Stillinger, Phys. Rev. E,
{\bf 68}, 041113 (2003).
\bibitem {libro} A. Gabrielli, F. Sylos Labini, M. Joyce, and
L. Pietronero, {\em Statistical Physics for Cosmic Structures},
(Springer-Verlag Inc., Berlin, 2004).
\bibitem {gardiner} C. W. Gardiner, {\it Handbook of stochastic
 methods}, (Springer Verlag, Berlin, Second Edition, 1997).
\bibitem {bartlett} M. S. Bartlett, Jour. Roy. Stat. Soc., Ser. B
Methodol., {\bf 25}, 264 (1963).
\bibitem {random-levy} W. Paul and J. Baschnagel,
{\em Stochastic processes from physics to finance}
(Springer-Verlag, Berlin, 1999).
\bibitem {k4} A. Gabrielli, M. Joyce, B. Marcos, and P. Viot, 
Europhys. Lett., {\bf 66}, 1 (2004).
\bibitem {adler} R. J. Adler, {\em The Geometry of Random Fields},
(Wiley, London, 2001).
\bibitem {bart-th1} J. F. C. Kingman, {\em Poisson processes}, Oxford
Studies in Probability, vol. 3, Clarendon Press, Oxford University Press
(New York, 1993). MR 94a:60052
\bibitem {bart-th2} O. Knill, Electronic Res. Announcements of
the Am. Math. Soc., {\bf 3}, 110 (1997).


\end{thebibliography}
\end{document}